\documentclass[vecphys]{svmult}

\usepackage{graphicx}        
\usepackage{multicol}        
\usepackage[bottom]{footmisc}

\newcommand{\la}{\left\langle}
\newcommand{\ra}{\right\rangle}
\newcommand{\EPL}{Europhys.~Lett.~}
\newcommand{\PRL}{Phys.~Rev.~Lett.~}
\newcommand{\PR}{Phys.~Rev.~}
\newcommand{\JCP}{J.~Chem.~Phys.~}

\newcommand{\JSP}{J.~Stat.~Phys.~}
\newcommand{\JPCM}{J.~Phys.: Condens.~Matter~}
\newcommand{\MP}{Mol.~Phys.~}
\newcommand{\JCIS}{J.~Coll.~Int.~Sci.~}

\begin{document}

\title*{Effective Interactions in Soft Materials}
\author{Alan R. Denton}
\institute{Department of Physics, North Dakota State University,
Fargo, North Dakota, 58105-5566, U.S.A.~~\texttt{alan.denton@ndsu.edu}}
%
%
\maketitle


\section{Introduction}\label{Intro}

Soft condensed matter systems are typically multicomponent mixtures of 
macromolecules and simpler components that form complex structures spanning 
wide ranges of length and time scales~\cite{deGennes96,Witten99,Cates,
Hamley,Jones,Witten-Pincus}.  Common classes of soft materials are
colloidal dispersions~\cite{Hunter,Pusey,Schmitz,Evans}, 
polymer solutions and melts~\cite{deGennes79,Doi,Oosawa,Hara}, 
amphiphilic systems~\cite{Gompper-Schick,Israelachvili94},
and liquid crystals~\cite{Frenkel,Chandrasekhar,deGennes93}.  
Among these classes are many biologically important systems, 
such as DNA, proteins, and cell membranes.  
Most soft materials are intrinsically nanostructured, in that at least 
some components display significant variation in structure on length scales 
of 1-10 nanometers.  
Many characteristic traits of soft condensed matter, e.g., mechanical 
fragility, sensitivity to external perturbation, and tunable thermal 
and optical properties, result naturally from the mingling of microscopic 
and mesoscopic constituents.

The complexity of composition that underlies the rich physical properties of 
soft materials poses a formidable challenge to theoretical and computational
modelling efforts.  Large size and charge asymmetries between macromolecules
(e.g., colloidal or polyelectrolyte macroions) and microscopic components
(e.g., counterions, monomers, solvent molecules) often render impractical 
the explicit modelling of all degrees of freedom over physically significant 
length and time scales.  Model complexity can be greatly reduced,
however, by pre-averaging (coarse-graining) the degrees of freedom 
of some of the microscopic components, thus mapping the original model 
onto an effective model, with a reduced number of components, governed by 
\textit{effective} interparticle interactions.

The concept of effective interactions has a long history in the statistical 
mechanics of liquids~\cite{HM} and other condensed matter systems, dating 
back over a half-century to the McMillan-Mayer theory of solutions~\cite{MM},
the Derjaguin-Landau-Verwey-Overbeek (DLVO) theory of charge-stabilised
colloids~\cite{DL,VO}, and the pseudopotential theory of simple metals and 
alloys~\cite{AS,Hafner}.  
In modelling materials properties of simple atomic or molecular liquids 
and crystals, it is often justifiable 
to average over electronic degrees of freedom of the constituent molecules.  
The coarse-grained model then comprises a collection of structureless 
particles interacting via effective intermolecular potentials whose 
parameters depend implicitly on the finer electronic structure.  In recent 
years, analogous methods have been carried over and adapted to the realm 
of macromolecular (soft) materials.

Effective interparticle interactions prove especially valuable in modelling
materials properties of soft matter systems, which depend on the collective 
behaviour of many interacting particles.  In studies of thermodynamic phase 
behaviour, for example, effective interactions provide essential input to 
molecular (e.g., Monte Carlo and molecular dynamics) simulations and 
statistical mechanical theories.  While such methods can be directly 
applied, in principle, to an explicit model of the system, brute force 
applications are, in practice, often simply beyond computational reach.  
Consider, for example, that a molecular simulation of only 1000 macroions,
each accompanied by as few as 100 counterions, entails following the motions 
of $10^5$ particles, computing at each step electrostatic and excluded-volume 
interactions among the particles.  A far more practical strategy applies
statistical mechanical methods to an effective model of fewer components.  

This chapter reviews the statistical mechanical foundations underlying theories 
of effective interparticle interactions in soft matter systems.  After first 
identifying and defining the main systems of interest in Sec.~\ref{Systems},
several of the more common theoretical methods are sketched in 
Sec.~\ref{Interactions}.  Concise derivations are given, in particular, for
response theory, density-functional theory, and distribution function theory.
Although these methods are all well established, the interconnections among 
them are not widely recognized.  An effort is made, therefore, to demonstrate 
the underlying unity of these seemingly disparate approaches.  
Practical implementations are illustrated in Sec.~\ref{Applications},
where recent applications to charged colloids, colloid-polymer mixtures, 
and polymer solutions are outlined.
In the limited space available, little more than a sample of many methods 
and applications can be included.  Complementary perspectives and details 
can be found in several excellent 
reviews~\cite{Hansen-Lowen,Belloni00,Likos01,Levin02}.  
Finally, in Sec.~\ref{Outlook}, a gaze into the (liquid) crystal ball 
portends an exciting outlook for the field.

\section{Systems of Interest}\label{Systems}

The main focus of this chapter is effective interactions among macromolecules 
dispersed in simple molecular solvents.  The term ``simple" here implies 
simplicity of molecular structure, not necessarily properties, thereby 
including water -- the most ubiquitous, biologically relevant, and anomalous 
solvent.  The macromolecules may be colloidal (or nano-) particles, polymers, 
or amphiphiles and may be electrically neutral or charged, as in the cases of 
charge-stabilised colloids, polyelectrolytes (including biopolymers), 
and ionic surfactants.  

Colloidal suspensions~\cite{Hunter} consist of ultra-divided matter 
dispersed in a molecular solvent, and are often classified as 
lyophobic (``solvent hating") or lyophilic (``solvent loving"), according 
to the ease with which the particles can be redispersed if dried out.
Depending on density, colloidal particles are typically nanometers to 
microns in size -- sufficiently large to exhibit random Brownian motion, 
perpetuated by collisions with solvent molecules, yet small enough to 
remain indefinitely suspended against sedimentation.  The upper size limit 
can be estimated by comparing the change in gravitational energy as a 
colloid traverses one particle diameter to the typical thermal energy 
$k_BT$ at absolute temperature $T$, where $k_B$ is Boltzmann's constant.
To better appreciate these length scales, consider repeatedly dividing 
a cube of side length 1 cm until reaching first the width of a human hair 
(10-100 $\mu$m), then the diameter of a colloid, and finally the diameter
of an atom.  How many cuts are required?

Polymers~\cite{deGennes79,Doi} are giant chainlike, branched, or networked 
molecules, consisting of covalently linked repeat units (monomers), which 
may be all alike (homopolymers) or of differing types (heteropolymers).  
Polyelectrolytes \cite{Oosawa,Hara} are polymers that carry ionizable groups.
Depending on the nature of intramolecular monomer-monomer interactions, 
polymer and polyelectrolyte chains may be stiff or flexible.  Flexibility 
can be quantified by defining a persistence length as the correlation length 
for bond orientations, i.e., the distance along the chain over which bond 
orientations become decorrelated.  At one extreme, rigid rodlike polymers 
have a persistence length equal to the contour length of the chain.  
At the opposite extreme, freely-jointed polymers are random-walk coils with
spatial extent best characterized by the radius of gyration, defined as 
the root-mean-square displacement of monomers from the chain's centre of mass.

Amphiphilic molecules~\cite{Gompper-Schick,Israelachvili94} consist of a 
hydrophilic head group joined to a hydrophobic tail group, usually a 
hydrocarbon chain.  The head group may be charged (ionic) or neutral 
(nonionic).  When sufficiently concentrated in aqueous solution, surfactants 
and other amphiphiles organize (self-assemble) into regular structures 
that optimize exposure of head groups to the exterior water phase, while 
sequestering the hydrophobic tails within.  The various structures include 
spherical and cylindrical micelles, bilayers, vesicles (bilayer capsules), 
and microemulsions.  Common soap films, for example, are 
bilayers of surfactants (surfactant-water-surfactant sandwiches) immersed 
in air, while biological membranes are bilayers of phospholipids immersed
in water.  Relative stabilities of competing structures are governed largely
by concentration and geometric packing constraints, as determined by the 
relative sizes of head and tail groups~\cite{Israelachvili85}.

Colloids, polyelectrolytes, and amphiphiles can acquire charge in solution 
through dissociation of ions from chemical groups on the colloidal surfaces, 
polymer backbones, or amphiphile head groups.  For ions of sufficiently 
low valence, the entropic gain upon dissociation exceeds the energetic cost 
of charge separation, resulting in a dispersion of charged macroions and
an entourage of oppositely charged counterions.  In an electrolyte solvent, 
charged macroions interact with one another, and with charged surfaces, 
via electrostatic interactions that are screened by intervening counterions 
and salt ions in solution.  Equilibrium and nonequilibrium distributions of 
ions are determined by a competition between entropy and various microscopic
interactions~\cite{Israelachvili85}, including repulsive Coulomb and steric 
interactions and attractive van der Waals (e.g., dipole-induced-dipole) 
interactions.  By changing system parameters, such as macroion properties 
(size, charge, composition), and solvent properties (salt concentration, pH, 
dielectric constant, temperature), the range and strength of interparticle 
interactions can be widely tuned.  Rational control over the enormously 
rich equilibrium and dynamical properties of macromolecular materials 
relies on a fundamental understanding of the nature and interplay of 
microscopic interactions.

In all of the systems of interest, the macromolecules possess some degree
of internal structure.  In charged colloids and polyelectrolytes, a multitude 
of microscopic degrees of freedom are associated with the distribution of 
charge over the macroion and the distribution of counterions throughout the 
solvent.  In polymer and amphiphilic solutions, the polymer chains or 
amphiphilic assemblies have conformational freedom.  Furthermore, the solvent 
itself contributes a vast number of molecular degrees of freedom.  
The daunting prospect of explicitly modelling multicomponent mixtures on a 
level so fine as to include all molecular degrees of freedom motivates the 
introduction of effective models governed by effective interactions.  
The loss of structural information upon coarse graining necessitates an
inevitable compromise in accuracy.  The art of deriving and applying
effective interactions lies in crafting approximations that are 
computationally manageable yet capture the essential physics.

\section{Effective Interaction Methods}\label{Interactions}

\subsection{Statistical Mechanical Foundation}\label{Statmech}

Effective interaction methods have a rigorous foundation in the statistical 
mechanics of mixtures~\cite{Rowlinson84}.  These methods rest on the premise
that, by averaging over the degrees of freedom of some of the components,
a multicomponent mixture can be mapped onto an effective model, with a 
reduced number of components.  While the true mixture is subject to 
bare interparticle interactions, the reduced model is governed by 
coarse-grained, effective interactions.  In charged colloids, for example, 
averaging over coordinates of the solvent molecules and of the microions 
(counterions and salt ions) maps the suspension onto an effective 
one-component model of mesoscopic ``pseudomacroions" subject to 
microion-induced effective interactions.  The bare electrostatic (Coulomb) 
interactions between macroions in the true suspension are replaced by 
screened-Coulomb interactions in the effective model. 
Similarly, in mixtures of colloids and non-adsorbing polymers, averaging 
over polymer degrees of freedom leads to polymer-induced effective 
interactions between the colloids.

\subsubsection{Generic Two-Component Model}\label{Model1}

We consider here a simple, pairwise-interacting, two-component mixture of $N_a$ 
particles of type $a$ and $N_b$ of type $b$, obeying classical statistics, 
confined to volume $V$ at temperature $T$.  A given system is modelled by 
three bare interparticle pair potentials, $v_{aa}(r)$, $v_{bb}(r)$, and 
$v_{ab}(r)$, assumed here to be isotropic, where $r$ is the distance between 
the particle centres.  The model system could represent, e.g., 
a charge-stabilised colloidal suspension, 
a polyelectrolyte solution, or various mixtures of colloids, nonadsorbing 
polymers, nanoparticles, or amphiphilic assemblies (micelles, vesicles, etc.).
For simplicity, the discussion is confined to binary mixtures, although
the methods discussed below easily generalize to multicomponent mixtures.
Throughout the derivations, it may help to visualize, for concreteness,
the $a$ particles as colloids and the $b$ particles as counterions.
\begin{figure}
\centering
\includegraphics[height=4cm]{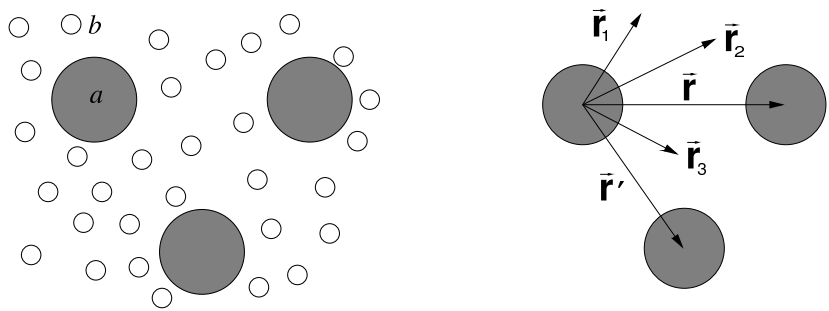}
\caption{\label{fig-diagram} 
Left: Generic model of a binary mixture of species labelled $a$ and $b$.
Right: Effective one-component model, after coarse-graining of $b$ species,
and geometry for physical interpretation 
of response theory.  Vectors ${\bf r}$ and ${\bf r}'$ define 
centre-to-centre displacements of $a$ particles.
Vectors ${\bf r}_1$, ${\bf r}_2$, and ${\bf r}_3$ define points at which 
either the ``external" potential of the $a$ particles acts or a change 
is induced in the density of $b$ particles (see Sec.~\ref{Interpretation}).}
\end{figure}

Even before analysing the system in detail, it should be conceptually apparent 
that, since particles inevitably influence their environment, the presence of 
particles of one species can affect the manner in which all other particles 
interact.  Familiar analogies may be identified in any mixture of interacting
entities -- from inanimate particles to living cells, organisms, and ecosystems.
In a simple binary mixture, for example, particles of type $b$ can be regarded 
as \textit{inducing} interactions between $a$ particles.  The induced 
interactions, which act in addition to bare $aa$ interactions, depend on both 
the $ab$ interactions and the distribution of $b$ particles.  The effective 
interactions, which are simply sums of induced and bare interactions, may be 
many-body in character, even if all bare interactions are strictly pairwise, 
and may depend on the thermodynamic state (density, temperature, etc.) 
of the system.

\subsubsection{Reduction to Effective One-Component Model}\label{Reduction}

These qualitative observations are now quantified by developing a statistical 
description of the system.  We start from the Hamiltonian function $H$, which 
governs all equilibrium and dynamical properties of the system, and assume 
pairwise bare interactions.  The Hamiltonian naturally separates, according to 
$H=K+H_{aa}+H_{bb}+H_{ab}$, into the kinetic energy $K$ and three interaction 
terms:
\begin{equation}
H_{\alpha\alpha}=\frac{1}{2}\sum_{{i\neq j=1}}^{N_{\alpha}}
v_{\alpha\alpha}(r_{ij}), \qquad \alpha=a,b
\label{Ha1}
\end{equation}
and
\begin{equation}
H_{ab}=\sum_{{i=1}}^{N_a}\sum_{{j=1}}^{N_b} v_{ab}(r_{ij}),
\label{Hab1}
\end{equation}
where $r_{ij}=|{\bf r}_i-{\bf r}_j|$ denotes the separation between the 
centres of particles $i$ and $j$, at positions ${\bf r}_i$ and ${\bf r}_j$.
Within the canonical ensemble (constant $N_a$, $N_b$, $V$, $T$), the 
thermodynamic behaviour of the system is governed by the canonical 
partition function
\begin{equation}
{\cal Z}=\la\la\exp(-\beta H)\ra_a\ra_b, 
\label{part}
\end{equation}
where $\beta=1/k_BT$ and $\la\cdots\ra_{\alpha}$ denotes a classical 
canonical trace over the coordinates of particles of type $\alpha$:
\begin{equation}
\la\exp(-\beta H)\ra_{\alpha}=\frac{1}{N_{\alpha}!
\Lambda_{\alpha}^{3N_{\alpha}}}\int{\rm d}{\bf r}_1
\cdots\int{\rm d}{\bf r}_{N_{\alpha}}\,\exp(-\beta H),
\label{trace}
\end{equation}
with $\Lambda_{\alpha}$ being the respective thermal de Broglie wavelength.

The two-component mixture can be formally mapped onto an equivalent 
one-component system by performing a restricted trace over the coordinates 
of only the $b$ particles, keeping the $a$ particles fixed.  
Thus, without approximation,
\begin{equation}
{\cal Z}=\la\exp(-\beta H_{aa})\la\exp[-\beta(H_{bb}+H_{ab})]\ra_b\ra_a
=\la\exp(-\beta H_{\rm eff})\ra_a, 
\label{parta}
\end{equation}
where 
\begin{equation}
H_{\rm eff}=H_{aa}+F_b 
\label{Heff}
\end{equation}
is the effective Hamiltonian of the equivalent one-component system and 
\begin{equation}
F_b=-k_BT\ln\la\exp\left[-\beta(H_{bb}+H_{ab})\right]\ra_b
\label{Fb1}
\end{equation}
can be physically interpreted as the Helmholtz free energy of the $b$ particles 
in the presence of the fixed $a$ particles.  
Equations~(\ref{parta})-(\ref{Fb1}) provide a formally exact basis for 
calculating the effective interactions.  It remains, in practice, to explicitly
determine the effective Hamiltonian by approximating the ensemble average in 
(\ref{Fb1}).  Next, we describe three general methods of attack -- response 
theory, density-functional theory, and distribution function theory.  While 
these methods ultimately give equivalent results, they have somewhat differing 
origins and conceptual interpretations.

\subsection{Response Theory}\label{Response}

The term ``response theory" can have varying specific meanings, depending on 
discipline and context, but is used here to denote a collection of statistical
mechanical methods that describe the response, in a multicomponent condensed 
matter mixture, of the density of one component to the potential generated 
and imposed by another component.  Response theory has been systematically
developed and widely applied, over the past four decades, in the theory of 
simple metals~\cite{HM,AS,Hafner} to describe the quantum mechanical response
of valence electron density to the electrostatic potential of metallic ions.  
More recently, similar methods have been carried over and adapted to 
classical soft matter systems, in particular, to charge-stabilised colloidal 
suspensions~\cite{Silbert91,Denton99,Denton00}, 
polyelectrolytes~\cite{Denton03,Wang-Denton04,Wang-Denton05}, and 
colloid-polymer mixtures~\cite{Dijkstra-jpcm99,Dijkstra-pre99}.  Although 
most applications have been restricted to the linear response approximation, 
which assumes a linear dependence between the imposed potential (cause) and 
the density response (effect), increasing attention is being devoted to 
nonlinear response.  Below, we outline the key elements of response theory, 
including both linear and nonlinear approximations, in the context of a 
classical binary mixture.

\subsubsection{Perturbation Theory}\label{Perturbation}

To approximate the free energy (\ref{Fb1}) of one component 
($b$ particles) in the presence of another component ($a$ particles) 
it is often constructive to regard the $a$ particles as generating an 
``external" potential that perturbs the $b$ particles and induces their 
response.  This external potential, which depends on the $ab$ interaction 
and on the number density $\rho_a({\bf r})$ of $a$ particles, 
can be expressed as 
\begin{equation}
v_{\rm ext}({\bf r})=\int{\rm d}{\bf r}'\, v_{ab}(|{\bf r}-{\bf r}'|)
\rho_a({\bf r}'),
\label{vext}
\end{equation}
and the $ab$ interaction term in the Hamiltonian as
\begin{equation}
H_{ab}=\int{\rm d}{\bf r}\, \rho_b({\bf r}) v_{\rm ext}({\bf r}),
\label{Hab2}
\end{equation}
where 
\begin{equation}
\rho_{\alpha}({\bf r})=\sum_{i=1}^{N_{\alpha}}\delta({\bf r}-{\bf r}_i) 
\label{rho_alpha}
\end{equation}
is the number density operator for particles of type $\alpha$ ($\alpha=a,b$). 

If the $a$ particles possess a property (e.g., electric charge or size)
that can be continuously varied to tune the strength of $v_{\rm ext}({\bf r})$,
then $F_b$ can be approximated via a perturbative response theory.
A prerequisite for this approach is accurate knowledge of the free energy of 
a reference system of pure $b$ particles (unperturbed by the $a$ particles). 
Relative to this reference system, the free energy can be expressed as
\begin{equation}
F_b=F_0+\int_0^1{\rm d}\lambda\,\frac{\partial F_b(\lambda)}{\partial\lambda}
=F_0+\int_0^1{\rm d}\lambda\,\la H_{ab}\ra_{\lambda},
\label{Fb2}
\end{equation}
where 
\begin{equation}
F_0=F_b(\lambda=0)=-k_BT\ln\la\exp(-\beta H_{bb})\ra_b
\label{F0}
\end{equation}
is the reference free energy,
\begin{equation}
F_b(\lambda)=-k_BT\ln\la\exp\left[-\beta(H_{bb}+\lambda H_{ab})\right]\ra_b
\label{Fb-lambda}
\end{equation}
is the free energy of $b$ particles in the presence of $a$ particles 
``charged" to a fraction $\lambda$ of their full strength, and 
\begin{equation}
\la H_{ab}\ra_{\lambda}=\frac{\la H_{ab}\exp\left[-\beta(H_{bb}+\lambda H_{ab})
\right]\ra_b}{\la\exp\left[-\beta(H_{bb}+\lambda H_{ab})\right]\ra_b}
=\frac{\partial F_b(\lambda)}{\partial\lambda}
\label{Hab3}
\end{equation}
denotes a trace of $H_{ab}$ over the coordinates of the $b$ particles in this 
intermediate ensemble.
(To simplify notation, we henceforth omit the subscript $b$ from the trace
over the coordinates of $b$ particles: $\la\cdots\ra_b\equiv\la\cdots\ra$.)

Applying now a standard perturbative approximation, adapted from the theory 
of simple metals~\cite{HM,AS,Hafner}, the ensemble-averaged 
induced density of $b$ particles may be expanded in a functional Taylor series 
around the reference system [$v_{\rm ext}({\bf r})=0$] in powers of the 
dimensionless potential $u({\bf r})=-\beta v_{\rm ext}({\bf r})$:
\begin{equation}
\la\rho_b({\bf r})\ra=n_b+\sum_{n=1}^{\infty}\frac{1}{n!}\int{\rm d}{\bf r}_1\,
\cdots\int{\rm d}{\bf r}_n\, 
G^{(n+1)}({\bf r}-{\bf r}_1,\ldots,{\bf r}-{\bf r}_n)
u({\bf r}_1)\cdots u({\bf r}_n),
\label{delta-rhob}
\end{equation}
where $n_b=N_b/V$ is the average density of $b$ particles and the coefficients
\begin{equation}
G^{(n+1)}({\bf r}-{\bf r}_1,\ldots,{\bf r}-{\bf r}_n)=\lim_{u\to 0}
\left(\frac{\delta^n\la\rho_b({\bf r})\ra}
{\delta u({\bf r}_1)\cdots\delta u({\bf r}_n)}\right)
\label{G}
\end{equation}
are the $(n+1)$-particle density correlation functions~\cite{HM} of the 
reference system.  Equation~(\ref{delta-rhob}) has a simple physical 
interpretation: the density of $b$ particles induced at any point ${\bf r}$ 
results from the cumulative response to the external potentials at all 
points $\{{\bf r}_1,\ldots,{\bf r}_n\}$, propagated through the system 
via multiparticle density correlations.

Further progress follows more rapidly in Fourier space, where the 
self-interaction terms in the Hamiltonian (\ref{Ha1}) can be expressed 
using the identity
\begin{eqnarray}
\sum_{{i\neq j=1}}^{N_{\alpha}} v_{\alpha\alpha}(|{\bf r}_i-{\bf r}_j|)
&=&\int{\rm d}{\bf r}\int{\rm d}{\bf r}'\,\rho_{\alpha}({\bf r})
\rho_{\alpha}({\bf r}')v_{\alpha\alpha}(|{\bf r}-{\bf r}'|)
-N_{\alpha}v_{\alpha\alpha}(0) \nonumber \\
&=&\frac{1}{V}\sum_{\bf k}\hat v_{\alpha\alpha}(k)\left[
\hat\rho_{\alpha}({\bf k})\hat\rho_{\alpha}(-{\bf k})-N_{\alpha}\right],
\label{Hidentity1}
\end{eqnarray}
while the cross-interaction term (\ref{Hab1}) takes the form
\begin{equation}
\la H_{ab}\ra_{\lambda}=\frac{1}{V}\sum_{\bf k} \hat v_{ab}(k) 
\hat\rho_a({\bf k}) \la\hat\rho_b(-{\bf k})\ra_{\lambda}.
\label{Hab4}
\end{equation}
Here $\hat v_{\alpha\beta}(k)$ ($\alpha,\beta=a,b$) is the Fourier transform 
of the pair potential $v_{\alpha\beta}(r)$ and
\begin{equation}
\hat\rho_{\alpha}({\bf k})=\int{\rm d}{\bf r}\,\rho_{\alpha}({\bf r})
e^{-i{\bf k}\cdot{\bf r}}
\label{FTa}
\end{equation}
is the Fourier transform of the number density operator
(\ref{rho_alpha}), with inverse transform 
\begin{equation}
\rho_{\alpha}({\bf r})=\frac{1}{V}\sum_{\bf k}\hat\rho_{\alpha}({\bf k})
e^{i{\bf k}\cdot{\bf r}}.
\label{FTb}
\end{equation}
The inverse transform is expressed as a summation, rather than as an integral,
to allow the possibility of isolating the $k=0$ component to preserve the 
constraint of fixed average density in the canonical ensemble: 
$\hat\rho_{\alpha}(k=0)=\int{\rm d}{\bf r}\,\rho_{\alpha}({\bf r})=N_{\alpha}$.
For charged systems, which interact via bare Coulomb pair potentials, special care 
must be taken to ensure that all long-wavelength divergences formally cancel 
(see Sec.~\ref{Colloids} below).

Now Fourier transforming (\ref{delta-rhob}), we obtain 
\begin{equation}
\la\hat\rho_b({\bf k})\ra=\hat G^{(2)}(k)\hat u({\bf k})
+\frac{1}{2V}\sum_{{\bf k}'}\hat G^{(3)}({\bf k}',{\bf k}
-{\bf k}')\hat u({\bf k}')\hat u({\bf k}-{\bf k}')+\cdots,
\qquad k\neq 0,
\label{rhobk1}
\end{equation}
where the coefficients $\hat G^{(n)}$ (Fourier transforms of $G^{(n)}$) are 
related to the $n$-particle static structure factors of the reference system
via $\hat G^{(n)}=n_b S^{(n)}$, with the static structure factors being 
explicitly defined by~\cite{HM}
\begin{equation}
S^{(2)}(k)\equiv S(k)
=\frac{1}{N_b}\la\hat\rho_b({\bf k})\hat\rho_b(-{\bf k})\ra
\label{S2}
\end{equation}
and
\begin{equation}
S^{(n)}({\bf k}_1,\cdots,{\bf k}_{n-1})=\frac{1}{N_b}\la\hat\rho_b({\bf k}_1)
\cdots\hat\rho_b({\bf k}_{n-1})\hat\rho_b(-{\bf k}_1-\dots-{\bf k}_{n-1})\ra,
\quad n\geq 3.
\label{Sn}
\end{equation}
Substituting $\hat u({\bf k})=-\beta\hat v_{ab}(k)\hat\rho_a({\bf k})$
[from (\ref{vext})] into (\ref{rhobk1}), 
the induced density of $b$ particles can be expressed in the equivalent form
\begin{eqnarray}
\la\hat\rho_b({\bf k})\ra&=&\chi(k)
\hat v_{ab}(k)\hat\rho_a({\bf k})+\frac{1}{V}\sum_{{\bf k}'}
\chi'({\bf k}',{\bf k}-{\bf k}')
\hat v_{ab}(k') \hat v_{ab}(|{\bf k}-{\bf k}'|) \nonumber \\
&\times&\hat\rho_a({\bf k}') \hat\rho_a({\bf k}-{\bf k}')+\cdots, 
\qquad k\neq 0,
\label{rhobk2}
\end{eqnarray}
where 
\begin{equation}
\chi(k)=-\beta n_b S(k)
\label{chi1k}
\end{equation}
is the linear response function and
\begin{equation}
\chi'({\bf k}',{\bf k}-{\bf k}')=\frac{1}{2}\beta^2n_b
S^{(3)}({\bf k}',{\bf k}-{\bf k}') 
\label{chi2k}
\end{equation}
is the first nonlinear response function of the reference system.

Substituting the equilibrium density of $b$ particles (\ref{rhobk2}) 
into the $ab$ interaction (\ref{Hab4}), the latter into the free energy 
(\ref{Fb2}), and integrating over $\lambda$, yields the desired 
free energy of the $b$ particles to third order in the macroion density:
\begin{eqnarray}
F_b&=&F_0+N_a n_b\lim_{k\to 0}\hat v_{ab}(k)+
\frac{1}{2V}\sum_{{\bf k}\neq 0}\chi(k)\left[\hat v_{ab}(k)\right]^2
\hat\rho_a({\bf k})\hat\rho_a(-{\bf k}) 
\nonumber \\
&+&\frac{1}{3V^2}\sum_{{\bf k}\neq 0}\sum_{{\bf k}'}\chi'({\bf k}',-{\bf k}-{\bf k}') 
\hat v_{ab}(k)\hat v_{ab}(k')\hat v_{ab}(|{\bf k}+{\bf k}'|)
\nonumber \\
&\times&\hat\rho_a({\bf k})\hat\rho_a({\bf k}')\hat\rho_a(-{\bf k}-{\bf k}').
\label{Fb4}
\end{eqnarray}
Finally, this free energy may be substituted back into (\ref{Heff})
to obtain the effective Hamiltonian. 
Evidently, the term in $F_b$ that is quadratic in $\hat\rho_a({\bf k})$,
arising from the term in $\la\hat\rho_b({\bf k})\ra$ that is linear in 
$\hat\rho_a({\bf k})$, is connected to an effective interaction between 
pairs of $a$ particles.  Similarly, the term in $F_b$ that is cubic in 
$\hat\rho_a({\bf k})$, coming from the quadratic term in 
$\la\hat\rho_b({\bf k})\ra$, is connected to an effective interaction 
among triplets of $a$ particles.

\subsubsection{Effective Interparticle Interactions}\label{Effint}

To explicitly demonstrate the connections between the free energy $F_b$ 
and the effective interactions, we first identify
\begin{equation}
\hat v^{(2)}_{\rm ind}(k)=\chi(k)[\hat v_{ab}(k)]^2
\label{v2indk}
\end{equation}
as the interaction between pairs of $a$ particles induced by surrounding $b$ 
particles, in a linear response approximation~\cite{Silbert91,Denton99,Denton00}.
As expected, the induced interaction depends on both the bare $ab$ interaction 
and the response of the $b$ particles to the external potential of the $a$ 
particles.  Combining the bare $aa$ interaction with the induced interaction 
yields the linear response prediction for the effective pair interaction:
\begin{equation}
\hat v^{(2)}_{\rm lin}(k)=\hat v_{aa}(k)+\hat v^{(2)}_{\rm ind}(k).
\label{v2k}
\end{equation}
The term on the right side of (\ref{Fb4}) that is second-order 
in $\hat\rho_a({\bf k})$ can be manipulated using the identity 
[from (\ref{Hidentity1})]
\begin{equation}
\sum_{{i\neq j=1}}^{N_a} v^{(2)}_{\rm ind}(r_{ij})=\frac{1}{V}\sum_{\bf k\neq 0}
\hat v^{(2)}_{\rm ind}(k) \hat\rho_a({\bf k})\hat\rho_a(-{\bf k})
+N_a n_a\lim_{k\to 0}\hat v_{\rm ind}^{(2)}(k)-N_a v^{(2)}_{\rm ind}(0),
\label{Fb4-3}
\end{equation}
where $n_a=N_a/V$ is the average density of $a$ particles.
Similarly, identifying 
\begin{equation}
\hat v^{(3)}_{\rm eff}({\bf k},{\bf k}')=2\chi'({\bf k}',-{\bf k}-{\bf k}')
\hat v_{ab}(k)\hat v_{ab}(k')\hat v_{ab}(|{\bf k}+{\bf k}'|) 
\label{v3effk}
\end{equation}
in (\ref{Fb4}) as an effective three-body interaction, arising from 
nonlinear response, and invoking the identity
\begin{eqnarray}
\sum_{{i\neq j\neq k=1}}^{N_a}v^{(3)}_{\rm eff}({\bf r}_{ij},{\bf r}_{ik})
&=&\frac{1}{V^2}\sum_{\bf k}\sum_{{\bf k}'}
\hat v^{(3)}_{\rm eff}({\bf k},{\bf k}')
[\hat\rho_a({\bf k})\hat\rho_a({\bf k}')\hat\rho_a(-{\bf k}-{\bf k}')
\nonumber \\
~&-&~3\hat\rho_a({\bf k})\hat\rho_a(-{\bf k})+2N_a],
\label{Fb4-4}
\end{eqnarray}
the effective Hamiltonian acquires the following physically intuitive 
structure:
\begin{equation}
H_{\rm eff}=E+\frac{1}{2}\sum_{i\neq j=1}^{N_a}
v^{(2)}_{\rm eff}(r_{ij})+\frac{1}{3!}\sum_{i\neq j\neq k=1}^{N_a} 
v^{(3)}_{\rm eff}({\bf r}_{ij},{\bf r}_{ik}),
\label{Heff2}
\end{equation}
where $E$, $v^{(2)}_{\rm eff}(r)$, and $v^{(3)}_{\rm eff}({\bf r},{\bf r}')$ 
are, respectively, a one-body ``volume energy" and effective pair and triplet 
interactions, induced by the $b$ particles, between the $a$ particles.  

A natural by-product of the reduction to an effective one-component system,
the volume energy is entirely independent of the $a$ particle positions.  
Collecting coordinate-independent terms, the volume energy, expressed as 
$E=E_{\rm lin}+\Delta E$, comprises a linear response approximation,
\begin{equation}
E_{\rm lin}=F_0+\frac{N_a}{2}v^{(2)}_{\rm ind}(0)+N_a\lim_{k\to 0}
\left[n_b\hat v_{ab}(k)-\frac{1}{2}n_a\hat v_{\rm ind}^{(2)}(k)\right],
\label{Elin}
\end{equation}
and nonlinear corrections, the first-order correction being
\begin{equation}
\Delta E=\frac{N_a}{6}\left[v^{(3)}_{\rm eff}(0,0)-\frac{n_a}{V}\sum_{\bf k}
\hat v_{\rm eff}^{(3)}({\bf k},0)\right].
\label{Delta-E}
\end{equation}

The effective pair interaction $v^{(2)}_{\rm eff}(r)$ in (\ref{Heff2}) 
is the transform of
\begin{equation}
\hat v^{(2)}_{\rm eff}(k)=\hat v^{(2)}_{\rm lin}(k)+
\Delta\hat v^{(2)}_{\rm eff}(k), 
\label{v2effk}
\end{equation}
where
\begin{equation}
\Delta\hat v^{(2)}_{\rm eff}(k)=\frac{1}{V}\sum_{{\bf k}'}\hat v^{(3)}_{\rm eff}
({\bf k},{\bf k}')-\frac{n_a}{3}\hat v_{\rm eff}^{(3)}({\bf k},0)
\label{Delta-v2effk}
\end{equation}
is the first nonlinear correction to the effective pair potential, while the 
effective triplet interaction $v^{(3)}_{\rm eff}({\bf r},{\bf r}')$ is the 
Fourier transform of (\ref{v3effk}).
Note that the final term on the right sides of (\ref{Elin}),
(\ref{Delta-E}), and (\ref{Delta-v2effk}) arise from the constraint of 
fixed average density.

Three observations are in order.  First, since the volume energy depends, 
in general, on the mean densities of both $a$ and $b$ particles, 
it contributes to the total free energy and, therefore, thermodynamics 
of the system.  This point has special significance in applications to 
charged systems, as discussed in Sec.~\ref{Colloids}.
Second, nonlinear response of the $b$ particles generates not only effective 
many-body interactions among the $a$ particles, but also corrections to both 
the effective pair interaction and the volume energy.  In fact, as is clear 
from (\ref{Delta-E}) and (\ref{Delta-v2effk}), the nonlinear corrections 
to $E$ and $v^{(2)}_{\rm eff}(r)$ are intimately related to many-body 
interactions.  Third, any influence of $bb$ interactions on the effective 
interactions enters through the free energy and response functions 
of the reference system.  Thus, the quality of the effective interactions is 
limited only by the accuracy to which the structure and thermodynamics of 
the pure $b$ fluid are known.

\subsubsection{Physical Interpretation}\label{Interpretation}

While response theory is most easily formulated in Fourier space, 
its physical interpretation is perhaps more transparent in real space.  
The induced pair interaction in the linear response approximation
(\ref{v2indk}) can be expressed in terms of real-space functions as
\begin{equation}
v^{(2)}_{\rm ind}(r)=\int{\rm d}{\bf r}_1\,\int{\rm d}{\bf r}_2\,
\chi(|{\bf r}_1-{\bf r}_2|)v_{ab}(r_1)v_{ab}(|{\bf r_2}-{\bf r}|).
\label{v2indr}
\end{equation}
Here $\chi(|{\bf r}_1-{\bf r}_2|)$ is the real-space linear response function,
which describes the change in the density of $b$ particles induced at point 
${\bf r}_2$ in response to an external potential applied at point ${\bf r}_1$.
Referring to Fig.~\ref{fig-diagram}, (\ref{v2indr}) can be interpreted as follows.
A particle of type $a$, centred at the origin in Fig.~\ref{fig-diagram}, 
generates an external potential $v_{ab}(r_1)$, which acts on $b$ particles
at all points ${\bf r}_1$.  This potential induces at point ${\bf r}_2$ 
a change in the density of $b$ particles given by 
$\int{\rm d}{\bf r}_1\, \chi(|{\bf r}_1-{\bf r}_2|) v_{ab}(r_1)$.  
This induced density, which depends on pair correlations (via $\chi$) in 
the intervening $b$ fluid, then interacts with a \textit{second} $a$ particle, 
at displacement ${\bf r}$ from the first.  The net result is an {\it effective} 
interaction between the pair of $a$ particles {\it induced} by the medium.
The linear response contribution to the volume energy (per particle)
associated with $ab$ interactions (\ref{Elin}) has a closely related form:
\begin{equation}
v^{(2)}_{\rm ind}(0)=\int{\rm d}{\bf r}_1\,\int{\rm d}{\bf r}_2\,
\chi(|{\bf r}_1-{\bf r}_2|)v_{ab}(r_1)v_{ab}(r_2).
\label{Er}
\end{equation}
The physical interpretation is similar, except that the induced density now 
interacts back with the first $a$ particle, generating a one-body (self) energy.
An analogous interpretation applies to nonlinear response and induced 
many-body interactions~\cite{Denton-pre04}.

\subsection{Density-Functional Theory}\label{DFT}

An alternative, yet ultimately equivalent, approach to deriving effective 
interactions, follows from classical density-functional theory 
(DFT)~\cite{Oxtoby,Evans92}.
Classical DFT has a long history, dating back a half-century to the 
earliest integral-equation theories of simple liquids~\cite{HM}. 
Following the establishment of formal foundations~\cite{Evans79}, DFT has been 
widely applied, in recent decades, to a variety of soft condensed matter 
systems, including colloids, polymers, and liquid crystals.  Connections 
between density-functional theory and effective interactions in charged 
colloids have been established by 
L\"owen {\it et al}~\cite{Lowen,Graf-Lowen} and 
van Roij {\it et al}~\cite{vRH,vRDH}.  The essence of the theory is most 
easily grasped in the context of an $ab$ mixture, where the challenge again 
is to approximate the free energy (\ref{Fb1}) of a fluid of $b$ particles 
in the presence of fixed $a$ particles.

The basis of the density-functional approach is the existence~\cite{Evans79} 
of a grand potential functional $\Omega_b[\rho_b]$ -- the square brackets
denoting a functional dependence -- with two essential properties:
$\Omega_b[\rho_b]$ is uniquely determined by the spatially-varying density 
$\rho_b({\bf r})$, for any given external potential $v_{\rm ext}({\bf r})$,
and is a minimum, at equilibrium, with respect to $\rho_b({\bf r})$.  
The grand potential functional is related to the Helmholtz free energy 
functional ${\cal F}_b[\rho_b]$ via the Legendre transform relation
\begin{equation}
\Omega_b[\rho_b]={\cal F}_b[\rho_b]-\mu_bN_b,
\label{Omegab}
\end{equation}
where $\mu_b$ is the chemical potential of $b$ particles.
The free energy functional naturally separates, according to
\begin{equation}
{\cal F}_b[\rho_b]={\cal F}_{\rm id}[\rho_b]+{\cal F}_{\rm ext}[\rho_b]
+{\cal F}_{\rm ex}[\rho_b],
\label{Fb-DFT1}
\end{equation}
into an ideal-gas term ${\cal F}_{\rm id}$, which is the free energy in the 
absence of any interactions, an ``external" term ${\cal F}_{\rm ext}$, which 
results from interactions with the external potential, and an ``excess" term 
${\cal F}_{\rm ex}$, due entirely to interparticle interactions.  
The purely entropic ideal-gas free energy is given exactly by 
\begin{equation}
{\cal F}_{\rm id}[\rho_b]=k_BT\int{\rm d}{\bf r}\, \rho_b({\bf r})
[\ln(\rho_b({\bf r})\Lambda_b^3)-1], 
\label{Fid1}
\end{equation}
where $\Lambda_b$ is the thermal wavelength of the $b$ particles, 
while the external free energy can be expressed as
\begin{equation}
{\cal F}_{\rm ext}[\rho_b]=\int{\rm d}{\bf r}\, \rho_b({\bf r})
v_{\rm ext}({\bf r}),
\label{Fext}
\end{equation}
which is equivalent to $H_{ab}$ in (\ref{Hab2}).
Inserting (\ref{Fid1}) and (\ref{Fext}) into (\ref{Fb-DFT1}) yields
\begin{equation}
{\cal F}_b[\rho_b]=k_BT\int{\rm d}{\bf r}\, \rho_b({\bf r})
[\ln(\rho_b({\bf r})\Lambda_b^3)-1]+\int{\rm d}{\bf r}\, \rho_b({\bf r})
v_{\rm ext}({\bf r})+{\cal F}_{\rm ex}[\rho_b].
\label{Fb-DFT2}
\end{equation}

The excess free energy can be expressed in the formally exact form,
\begin{eqnarray}
{\cal F}_{\rm ex}[\rho_b]&=&\frac{1}{2}\int{\rm d}{\bf r}\int{\rm d}{\bf r}'
\int_0^1{\rm d}\lambda\,
\rho_{bb}^{(2)}[\lambda v_{bb};{\bf r},{\bf r}']
v_{bb}(|{\bf r}-{\bf r}'|) \nonumber \\ 
&=&\frac{1}{2}\int{\rm d}{\bf r}\int{\rm d}{\bf r}'\rho_b({\bf r})
\rho_b({\bf r}')\int_0^1{\rm d}\lambda\,
g_{bb}^{(2)}[\lambda v_{bb};{\bf r},{\bf r}']v_{bb}(|{\bf r}-{\bf r}'|),
\label{Fex1}
\end{eqnarray}
where $\rho_{bb}^{(2)}[\lambda v_{bb};{\bf r},{\bf r}']$ is the two-particle 
number density (a unique functional of the pair potential), 
$g_{bb}^{(2)}[\lambda v_{bb};{\bf r},{\bf r}']$ is the corresponding pair 
distribution functional,
and $\lambda$ is a coupling (or charging) constant that ``turns on" the 
interparticle correlations.  In general, the pair distribution functional 
is not known exactly and must be approximated.  For weakly correlated
systems, it is often reasonable to adopt the mean-field approximation
\begin{equation}
{\cal F}_{\rm ex}[\rho_b]=\frac{1}{2}\int{\rm d}{\bf r}\,\int{\rm d}{\bf r}'\,
\rho_b({\bf r})\rho_b({\bf r}')v_{bb}(|{\bf r}-{\bf r}'|),
\label{Fex2}
\end{equation}
which amounts to entirely neglecting correlations and assuming 
$g_{bb}^{(2)}=1$.

In a further approximation, valid for weakly inhomogeneous densities, the 
ideal-gas free energy functional is expanded in a functional Taylor series 
around the average density, which is truncated at quadratic order:
\begin{eqnarray}
{\cal F}_{\rm id}[\rho_b]&\simeq&F_{\rm id}(n_b)+
\left(\frac{\delta{\cal F}_{\rm id}[\rho_b]}{\delta\rho_b({\bf r})}\right)_{n_b}
\int{\rm d}{\bf r}\, [\rho_b({\bf r})-n_b] \nonumber \\
&+&\frac{1}{2}\int{\rm d}{\bf r}\int{\rm d}{\bf r}'\,\left(\frac{\delta^2
{\cal F}_{\rm id}[\rho_b]}{\delta\rho_b({\bf r})\delta\rho_b({\bf r}')}\right)_{n_b}
[\rho_b({\bf r})-n_b][\rho_b({\bf r}')-n_b].
\label{Fid2}
\end{eqnarray}
The first term on the right,
$F_{\rm id}(n_b)=N_b k_BT[\ln(n_b\Lambda_b^3)-1]$, is the ideal-gas free energy
of a uniform fluid of $b$ particles.  The second (linear) term on the right 
vanishes identically by virtue of the constraint of constant average density.  
Evaluating the functional derivative in the of the third (quadratic) term,
\begin{equation}
\frac{\delta^2{\cal F}_{\rm id}[\rho_b]}{\delta\rho({\bf r})
\delta\rho({\bf r}')}
=\frac{k_BT}{\rho_b({\bf r})}\delta({\bf r}-{\bf r}'),
\label{coeff}
\end{equation}
and combining (\ref{Fb-DFT1}), (\ref{Fext}), (\ref{Fex2}), and (\ref{Fid2}), 
the mean-field free energy functional finally can be approximated by
\begin{eqnarray}
{\cal F}_b[\rho_b]&\simeq&N_b k_BT[\ln(n_b\Lambda_b^3)-1]+\frac{k_BT}{2n_b}
\int{\rm d}{\bf r}\,[\rho_b({\bf r})-n_b]^2 
+\int{\rm d}{\bf r}\,\rho_b({\bf r})v_{\rm ext}({\bf r}) \nonumber \\
&+&\frac{1}{2}\int{\rm d}{\bf r}\,\int{\rm d}{\bf r}'\,
\rho_b({\bf r})\rho_b({\bf r}')v_{bb}(|{\bf r}-{\bf r}'|).
\label{Fb-DFT3}
\end{eqnarray}

The {\it equilibrium} density is now determined by the minimization condition
\begin{equation}
\beta\frac{\delta\Omega_b[\rho_b]}{\delta\rho_b({\bf r})}=\ln(n_b\Lambda_b^3)+
\frac{\rho_b({\bf r})}{n_b}-1+\beta v_{\rm ext}({\bf r})+
\beta\int{\rm d}{\bf r}'\, \rho_b({\bf r}')v_{bb}(|{\bf r}-{\bf r}'|)
-\beta\mu_b=0.
\label{equil}
\end{equation}
Fourier transforming and solving for the equilibrium density yields
\begin{equation}
\la\hat\rho_b({\bf k})\ra=\frac{-\beta n_b\hat v_{\rm ext}({\bf k})}
{1+\beta n_b\hat v_{bb}(k)}=\chi(k)\hat v_{\rm ext}({\bf k}), \qquad k\neq 0,
\label{rhobk3}
\end{equation}
where 
\begin{equation}
\chi(k)=\frac{-\beta n_b}{1+\beta n_b\hat v_{bb}(k)}
\label{chik-DFT}
\end{equation}
is a mean-field approximation to the linear response function introduced 
above in (\ref{chi1k}).
Now expressing the free energy functional (\ref{Fb-DFT3}) 
in terms of Fourier components,
\begin{eqnarray}
{\cal F}_b[\rho_b]&\simeq&N_b k_BT[\ln(n_b\Lambda_b^3)-1]+
n_b\lim_{k\to 0}\left[\frac{1}{2}N_b\hat v_{bb}(k)+N_a\hat v_{ab}(k)\right]
\nonumber \\
&+&\frac{1}{V}\sum_{{\bf k}\neq 0}\hat\rho_b({\bf k})\hat v_{ab}(k)
\hat\rho_a({\bf k})
+\frac{1}{2V}\sum_{{\bf k}\neq 0}\left(\hat v_{bb}(k)+\frac{1}{\beta n_b}\right)
\hat\rho_b({\bf k})\hat\rho_b(-{\bf k}), \qquad
\label{Fb-DFT4}
\end{eqnarray}
and substituting for the equilibrium density from (\ref{rhobk3}), we obtain
-- to second order in the $a$ particle density -- the equilibrium Helmholtz 
free energy of $b$ particles in the presence of the fixed $a$ particles:
\begin{eqnarray}
F_b&=&N_b k_BT[\ln(n_b\Lambda_b^3)-1]+
n_b\lim_{k\to 0}\left[\frac{1}{2}N_b\hat v_{bb}(k)+N_a\hat v_{ab}(k)\right]
\nonumber \\
&+&\frac{1}{2V}\sum_{\bf k\neq 0}\chi(k)\left[\hat v_{ab}(k)\right]^2
\hat\rho_a({\bf k})\hat\rho_a(-{\bf k}).
\label{Fb-DFT5}
\end{eqnarray}
After identifying
\begin{equation}
F_0=N_b k_BT[\ln(n_b\Lambda_b^3)-1]+\frac{1}{2}N_b n_b\lim_{k\to 0}
\hat v_{bb}(k)
\label{F02}
\end{equation}
as the free energy of the uniform $b$ fluid in the absence of $a$ particles,
(\ref{Fb-DFT5}) is seen to have exactly the same form as (\ref{Fb4}),
to quadratic order in $\hat\rho_a({\bf k})$.
The same effective interactions thus result from linearized classical 
density-functional theory as from linear response theory.  Moreover, the same 
agreement is also found for the effective triplet interactions derived from 
nonlinear response theory~\cite{Denton-pre04} and nonlinear 
DFT~\cite{Lowen-Allahyarov}.

\subsection{Distribution Function Theory}\label{Distribution}

Still another statistical mechanical approach to calculating effective 
interactions is based on approximating equilibrium distribution functions.
This approach has been developed by many workers and applied to 
charged colloids in the forms of various integral-equation 
theories~\cite{Patey80,Belloni86,Khan87,Carbajal-Tinoco02,
Petris02,Anta,Outhwaite02} and an extended Debye-H\"uckel 
theory~\cite{Chan85,Chan-pre01,Chan-langmuir01,Warren}.
The basic elements of the method are sketched below, 
again in the context of a simple $ab$ mixture.

As shown above in Sec.~\ref{Statmech}, the key quantity in any theory of 
effective interactions is the trace over degrees of freedom of the $b$ 
particles of the $ab$ interaction term in the Hamiltonian 
[see (\ref{Hab1}) and (\ref{Fb2})].  
This partial trace can be expressed in the general form
\begin{equation}
\la H_{ab}\ra=\int{\rm d}{\bf r}\int{\rm d}{\bf r}'\,\la \rho_a({\bf r})
\rho_b({\bf r}')\ra v_{ab}(|{\bf r}-{\bf r}'|),
\label{Hab5}
\end{equation}
where $\la\cdots\ra$ denotes an ensemble average over the coordinates of the 
$b$ particles with the $a$ particles fixed.  The density of the fixed $a$ 
particles being unaffected by the partial trace, we can replace
$\la\rho_a({\bf r})\rho_b({\bf r}')\ra$ by 
$\rho_a({\bf r})\la\rho_b({\bf r}')\ra$ in (\ref{Hab5}):
\begin{equation}
\la H_{ab}\ra=n_b\int{\rm d}{\bf r}\int{\rm d}{\bf r}'\,\rho_a({\bf r})
g_{ab}^{(2)}({\bf r}-{\bf r}') v_{ab}(|{\bf r}-{\bf r}'|),
\label{Hab6}
\end{equation}
thus introducing the $ab$ pair distribution function $g_{ab}^{(2)}({\bf r})$, 
which is defined via $\la\rho_b({\bf r})\ra=n_b g_{ab}^{(2)}({\bf r})$.  
This pair distribution function is proportional to the probability of finding 
a $b$ particle at displacement ${\bf r}$ from a central $a$ particle.  
More precisely, given an $a$ particle at the origin, 
$n_b g_{ab}^{(2)}({\bf r}){\rm d}{\bf r}$ represents the average number of 
$b$ particles in a volume ${\rm d}{\bf r}$ at displacement ${\bf r}$.

Distribution function theory evidently shifts the challenge to determining 
the cross-species ($ab$) pair distribution function.  To this end, we consider 
an approximation scheme -- rooted in the theory of simple liquids~\cite{HM} -- 
that illustrates connections to response theory and density-functional theory.  
The starting point is the fundamental relation, valid for any nonuniform fluid, 
between the equilibrium density, an ``external" applied potential, and the 
(internal) direct correlation functions.  Minimization of the Helmholtz 
free energy functional (\ref{Fb-DFT1}) with respect to the density, 
at fixed average density, yields the Euler-Lagrange relation~\cite{Evans79}
\begin{equation}
\la\rho_b({\bf r})\ra=\frac{e^{\beta\mu_b}}{\Lambda_b^3}
\exp(-\beta v_{\rm ext}({\bf r})+c_b^{(1)}[\rho_b;{\bf r}]),
\label{rhobr1}
\end{equation}
where the one-particle direct correlation functional (DCF), defined as
\begin{equation}
c_b^{(1)}[\rho_b;{\bf r}]\equiv -\beta\frac{\delta{\cal F}_{\rm ex}[\rho_b]}
{\delta\rho_b({\bf r})}=-\beta\mu_{\rm ex}[\rho_b;{\bf r}],
\label{c1}
\end{equation}
is a unique functional of the density that is proportional to the excess chemical 
potential of $b$ particles $\mu_{\rm ex}$ (associated with $bb$ 
interparticle interactions).

Although it provides an exact implicit relation for the equilibrium density,
(\ref{rhobr1}) can be solved, in practice, only by approximating 
$c_b^{(1)}[\rho_b;{\bf r}]$.  Approximations are facilitated by
expanding $c_b^{(1)}[\rho_b;{\bf r}]$ in a functional Taylor series 
about the average (bulk) density $n_b$:
\begin{equation}
c_b^{(1)}[\rho_b;{\bf r}]=c_b^{(1)}(n_b)+\int{\rm d}{\bf r}'\,
c_{bb}^{(2)}({\bf r}-{\bf r}';n_b)[\la\rho_b({\bf r}')\ra-n_b]+\cdots,
\label{c1-series}
\end{equation}
where 
\begin{equation}
c_{bb}^{(2)}({\bf r}-{\bf r}';n_b)=\lim_{\rho_b({\bf r})\to n_b}
\left(\frac{\delta c_b^{(1)}[\rho_b;{\bf r}]}{\delta\rho_b({\bf r}')}\right)
\label{c2}
\end{equation}
is the two-particle DCF of the (reference) uniform fluid and, more generally,
\begin{equation}
c_{b\cdots b}^{(n)}[\rho_b;{\bf r}_1,\ldots,{\bf r}_n]\equiv
\frac{\delta^{n-1} c_b^{(1)}[\rho_b;{\bf r}_1]}
{\delta\rho_b({\bf r}_2)\cdots\delta\rho_b({\bf r}_n)}
\label{cn}
\end{equation}
is the $n$-particle DCF.  Note that higher-order terms in the series,
corresponding to multiparticle correlations, are nonlinear in the density. 

Substituting (\ref{c1-series}) into (\ref{rhobr1}), and identifying 
$(e^{\beta\mu_b}/\Lambda_b^3)\exp[c_b^{(1)}(n_b)]$ as the bulk density $n_b$,
the nonuniform equilibrium density can be expressed as
\begin{equation}
\la\rho_b({\bf r})\ra=n_b\exp\left(-\beta v_{\rm ext}({\bf r})+\int{\rm d}
{\bf r}'\, c_{bb}^{(2)}({\bf r}-{\bf r}';n_b)[\la\rho_b({\bf r}')\ra-n_b]
+\cdots \right).
\label{rhobr2}
\end{equation}
Various practical approximations for the DCFs correspond, within the framework 
of integral-equation theory, to different closures of the Ornstein-Zernike 
relation~\cite{HM},
\begin{equation}
h_{bb}^{(2)}({\bf r})=c^{(2)}_{bb}({\bf r})+n_b\int{\rm d}{\bf r}'\, 
c^{(2)}_{bb}({\bf r}-{\bf r}')h_{bb}^{(2)}({\bf r}'),
\label{OZ}
\end{equation}
which is an integral equation relating the two-particle DCF to the 
pair correlation function, $h_{bb}^{(2)}({\bf r})=g_{bb}^{(2)}({\bf r})-1$.

Truncating the functional expansion in (\ref{rhobr2}) and retaining only the 
term linear in density, thus neglecting multiparticle correlations in a 
mean-field approximation, is equivalent to the hypernetted-chain (HNC) 
approximation in integral-equation theory.
Making a second mean-field approximation by equating the two-particle DCF
$c_{bb}^{(2)}({\bf r};n_b)$ to its asymptotic limit~\cite{HM},
\begin{equation}
\lim_{r\to\infty}c_{bb}^{(2)}({\bf r};n_b)=-\beta v_{bb}(r),
\label{c2-MFA}
\end{equation}
thereby neglecting short-range correlations, (\ref{rhobr2}) reduces to
\begin{equation}
\la\rho_b({\bf r})\ra=n_b\exp\left(-\beta v_{\rm ext}({\bf r})
-\beta\int{\rm d}{\bf r}'\,v_{bb}^{(2)}(|{\bf r}-{\bf r}'|)
[\la\rho_b({\bf r}')\ra-n_b]\right),
\label{rhobr4}
\end{equation}
corresponding to the mean-spherical approximation (MSA) in integral-equation 
theory.  In passing, we note that (\ref{rhobr4}) also provides a basis 
for the Poisson-Boltzmann theory of charged colloids and polyelectrolytes,
if we identify 
\begin{equation}
q_b\psi({\bf r})\equiv v_{\rm ext}({\bf r})+\int{\rm d}{\bf r}'\,
v_{bb}^{(2)}(|{\bf r}-{\bf r}'|)[\la\rho_b({\bf r}')\ra-n_b]
\label{psi}
\end{equation}
as the mean-field electrostatic potential energy of a $b$ particle 
(charge $q_b$) brought from infinity, where the potential $\psi=0$, 
to a displacement ${\bf r}$ away from an $a$ particle, and combine this 
with Poisson's equation,
\begin{equation}
\nabla^2\psi({\bf r})=-\frac{1}{\epsilon}\sum_i q_i\rho_i({\bf r}),
\label{Poisson}
\end{equation}
where $\epsilon$ is the dielectric constant of the medium and the sum is over 
all species with charges $q_i$ and number densities $\rho_i({\bf r})$.

If we now make one further approximation by linearizing the exponential 
function, valid in the case of potential energies much lower than thermal 
energies, then (\ref{rhobr4}) becomes
\begin{equation}
\la\rho_b({\bf r})\ra=n_b\left(1-\beta v_{\rm ext}({\bf r})-
\beta\int{\rm d}{\bf r}'\,v_{bb}^{(2)}(|{\bf r}-{\bf r}'|)
[\la\rho_b({\bf r}')\ra-n_b]\right).
\label{rhobr5}
\end{equation}
Fourier transforming (\ref{rhobr5}) and solving for the equilibrium
density finally yields
\begin{equation}
\la\hat\rho_b({\bf k})\ra=\frac{-\beta n_b\hat v_{\rm ext}({\bf k})}
{1+\beta n_b\hat v_{bb}(k)}=\chi(k)\hat v_{\rm ext}({\bf k}), \qquad k\neq 0,
\label{rhobk4}
\end{equation}
which is identical in form to the linear response and DFT predictions 
[see (\ref{rhobk2}) and (\ref{rhobk3})] with the same linear response 
function $\chi(k)$ as before [see (\ref{chik-DFT})].

Alternatively, we may first linearize the exponential in (\ref{rhobr4}) 
and then exploit the Ornstein-Zernike relation (\ref{OZ}) to solve 
recursively for the equilibrium density, with the result 
\begin{eqnarray}
\la\rho_b({\bf r})\ra&=&n_b\left(1-\beta v_{\rm ext}({\bf r})+
\int{\rm d}{\bf r}'c_{bb}^{(2)}({\bf r}-{\bf r}')[-\beta n_b 
v_{\rm ext}({\bf r}')+\cdots]\right) \nonumber \\
&=&n_b-\beta n_b\int{\rm d}{\bf r}'[\delta({\bf r}')+n_b h_{bb}^{(2)}({\bf r}')]
v_{\rm ext}({\bf r}'),
\label{rhobr6}
\end{eqnarray}
where 
\begin{equation}
-\beta n_b[\delta({\bf r})+n_b h_{bb}^{(2)}({\bf r})]=\chi({\bf r})
\label{chir1}
\end{equation}
can be identified as the real-space linear response function. 
By Fourier transforming (\ref{rhobr6}) and (\ref{chir1}), we recover the 
linear response relation (\ref{rhobk4}) with a linear response function,
\begin{equation}
\chi(k)=-\beta n_b[1+n_b\hat h_{bb}^{(2)}(k)]=-\beta n_b S(k),
\label{chik1}
\end{equation}
defined precisely as originally in (\ref{chi1k}).
The connection to the linear response function in (\ref{chik-DFT}) is 
established via the Fourier transform of the Ornstein-Zernike relation 
(\ref{OZ}),
\begin{equation}
\hat h_{bb}^{(2)}(k)=\frac{\hat c_{bb}^{(2)}(k)}{1-n_b\hat c_{bb}^{(2)}(k)}
\simeq\frac{-\beta\hat v_{bb}^{(2)}(k)}{1+\beta n_b\hat v_{bb}^{(2)}(k)},
\label{hbbk}
\end{equation}
where we have assumed a mean-field (random phase) approximation for the 
two-particle DCF, $\hat c_{bb}^{(2)}(k)\simeq -\beta\hat v_{bb}^{(2)}(k)$ 
[cf (\ref{c2-MFA})].  Distribution function theory therefore 
predicts the same linear response of the $b$-particle density to the 
$a$-particle external potential -- and thus the same effective interactions 
-- as do both response theory and density-functional theory.

In closing this section, we emphasize that the effective interactions derived 
here in the canonical ensemble apply to experimental situations in which 
the particle densities are fixed.  In many experiments, however, the system 
may be in chemical equilibrium with a reservoir of particles, allowing
fluctuations in particle densities.  The appropriate ensemble then would be 
the semigrand or grand canonical ensemble for a reservoir containing, 
respectively, one or both species.  It is left as an exercise to derive 
the effective interactions in these other ensembles.  
(Hint: Consider carefully the appropriate reference system.)

\section{Applications}\label{Applications}

The preceding section describes several general, albeit formal, approaches 
to modelling effective interparticle interactions in soft matter systems.  
To illustrate the practical utility of some of the methods, the following 
section briefly outlines applications to three broad classes of system: 
(1) charge-stabilised colloidal suspensions, (2) colloid-polymer mixtures, 
and (3) polymer solutions.  These systems exhibit two of the most common
forms of microscopic interactions, namely, electrostatic and excluded-volume 
interactions.  
Further details can be found in \cite{Likos01,Denton99,Denton00,
Dijkstra-jpcm99,Dijkstra-pre99,Denton-pre04}.

\subsection{Charged Colloids}\label{Colloids}

\subsubsection{Primitive Model}\label{Model2}

\begin{figure}
\centering
\includegraphics[height=4cm]{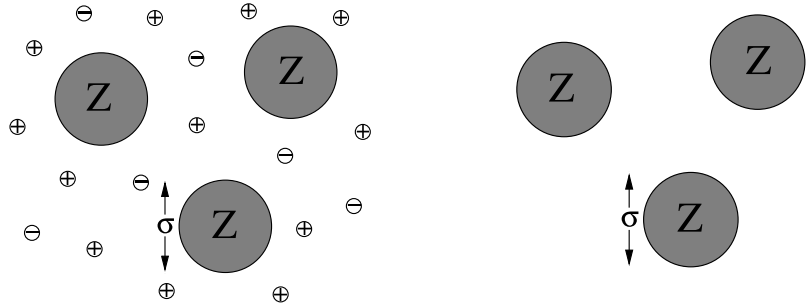}
\caption{\label{fig-model} Left: Primitive model of a charge-stabilised 
colloidal suspension, consisting of hard spherical macroions (valence $Z$, 
diameter $\sigma$) and point monovalent microions (counterions, salt ions) 
dispersed in a dielectric continuum.  Right: Effective one-component model 
of pseudomacroions, after coarse-graining of microions.}
\end{figure}

Charge-stabilised colloidal suspensions~\cite{Hunter,Pusey,Schmitz,Evans} are 
multicomponent mixtures of macroions, counterions, and salt ions dispersed 
in a molecular solvent and stabilised against coagulation by electrostatic
interparticle interactions.  For concreteness, we assume aqueous suspensions 
(water solvent).  A reasonable model for these complex systems is a 
collection of charged hard spheres and point microions interacting via 
bare Coulomb pair potentials in a dielectric medium (Fig.~\ref{fig-model}).  
The point-microion approximation is valid for systems, such as colloidal
suspensions, with large size asymmetries between macroions and microions.  
The macroions, of radius $a$ (diameter $\sigma=2a$), are assumed to carry a 
fixed, uniformly distributed, surface charge $-Ze$ (valence $-Z$), which may 
be physically interpreted as an effective charge, renormalized by association
of oppositely charged counterions (valence $z\ll Z$).  In a closed system,
described by the canonical ensemble, global charge neutrality constrains 
the macroion and counterion numbers, $N_m$ and $N_c$, via the relation 
$ZN_m=zN_c$.

The salt-water mixture is modelled as an electrolyte solution of $N_s$ 
dissociated pairs of point ions of valences $\pm z$.  The microions then number
$N_+=N_c+N_s$ positive and $N_-=N_s$ negative, totaling $N_{\mu}=N_c+2N_s$.  
Within the coarse-grained ``primitive" model of charged colloids, the water 
is treated as a dielectric continuum, characterized entirely by a dielectric 
constant $\epsilon$.  This approximation amounts to preaveraging over the 
vast number of solvent degrees of freedom.  For simplicity, we completely 
neglect charge-induced-dipole and other polarization 
interactions~\cite{Fisher94,Phillies95,Gonzalez01}, which are shorter-ranged 
than charge-charge interactions and vanish if solvent and macroions are 
index-matched (i.e., have the same dielectric constant).  The bare
electrostatic interactions can be represented by Coulomb pair potentials:
$v_{mm}(r)=Z^2e^2/\epsilon r$ ($r>\sigma$), $v_{cc}(r)=z^2e^2/\epsilon r$,
and $v_{mc}(r)=Zze^2/\epsilon r$ ($r>a$).  Note that in the primitive model, 
the solvent acts only to reduce the strength of Coulomb interactions 
by a factor $1/\epsilon$.  In addition, the macroion hard cores interact 
via a hard-sphere pair potential.

\subsubsection{Response Theory for Electrostatic Interactions}
\label{Response-EE}

Following the methods of response theory laid out in Sec.~\ref{Response}, 
effective interactions now can be derived by integrating out the degrees 
of freedom of the microions, reducing the multicomponent mixture to an 
effective one-component system of pseudo-macroions~\cite{Silbert91}.  
To simplify the derivation, we first consider the rather idealized case of 
salt-free suspensions.  The bare Hamiltonian of the two-component model 
then decomposes naturally, according to $H=H_{mm}+H_{cc}+H_{mc}$, into a 
macroion term
\begin{equation}
H_{mm}=H_{\rm HS}+\frac{1}{2}\sum_{{i\neq j=1}}^{N_m} v_{mm}(r_{ij}),
\label{Hm1}
\end{equation}
where $H_{\rm HS}$ is the Hamiltonian for neutral hard spheres 
(macroion hard cores), a counterion term
\begin{equation}
H_{cc}=K_c+\frac{1}{2}\sum_{{i\neq j=1}}^{N_c} v_{cc}(r_{ij}), 
\label{Hc1}
\end{equation}
where $K_c$ is the counterion kinetic energy, and a macroion-counterion 
interaction term
\begin{equation}
H_{mc}=\sum_{i=1}^{N_m}\sum_{j=1}^{N_c} v_{mc}(r_{ij}).
\label{Hmc1}
\end{equation}

By analogy with (\ref{Fb2}), the free energy of the counterions in the 
external potential of the macroions can be expressed as~\cite{HM,Silbert91}:
\begin{equation}
F_c=F_0+\int_0^1{\rm d}\lambda\,\la H_{mc}\ra_{\lambda},
\label{Fc2}
\end{equation}
where $F_0=-k_BT\ln\la\exp(-\beta H_c)\ra$ is now the reference free energy 
of the counterions in the presence of neutral (hard-core) macroions, and the 
$\lambda$-integral adiabatically charges the macroions from neutral to fully
charged.  Neglecting counterion structure induced by the macroion hard cores, 
neutral macroions would be surrounded by a uniform ``sea" of counterions.
As the macroion charge is turned on, the counterions respond, redistributing 
themselves to form a double layer (surface charge plus neighbouring counterions)
around each macroion.  

It is a special property of Coulomb-potential systems that, because volume 
integrals over long-ranged $1/r$ potentials diverge, each term on the right 
side of (\ref{Fc2}) is actually infinite.  Although the infinities 
formally cancel, it proves convenient still to convert $F_0$ to the free energy
of a classical one-component plasma (OCP) by adding and subtracting the 
(infinite) energy of a uniform compensating negative background
\begin{equation}
E_{bg}=-\frac{1}{2}N_cn_c\lim_{k\to 0}\hat v_{cc}(0), 
\label{Ebg}
\end{equation}
where $n_c$ is the average density of counterions in the volume unoccupied 
by the macroion cores.  Because the counterions are strictly excluded 
(with the background) from the hard macroion cores, the OCP has average density 
$n_c=N_c/[V(1-\eta)]$, where 
$\eta=\frac{\pi}{6}(N_m/V)\sigma^3$ is the macroion volume fraction and
$V(1-\eta)$ is the free volume.  Thus,
\begin{equation}
F_c=F_{\rm OCP}+\int_0^1{\rm d}\lambda\,\la H_{mc}\ra_{\lambda}-E_{bg},
\label{Fc3}
\end{equation}
where $F_{\rm OCP}=F_0+E_{bg}$ is the free energy of the ``Swiss cheese" OCP
in the presence of neutral, but volume-excluding, hard spheres.

All of the formal expressions derived in Sec.~\ref{Response} for a generic 
two-component ($ab$) mixture now carry over directly, with the identifications 
$a\leftrightarrow m$ and $b\leftrightarrow c$.  In the linear response 
approximation~\cite{Denton99,Denton00}, the volume energy is given by
\begin{equation}
E_{\rm lin}=F_{\rm OCP}+\frac{N_m}{2}v^{(2)}_{\rm ind}(0) 
+N_mn_c\lim_{k\to 0}\left[\hat v_{mc}(k)
-\frac{z}{2Z}\hat v^{(2)}_{\rm ind}(k)+\frac{Z}{2z}\hat v_{cc}(k)\right],
\label{Elin-cc}
\end{equation}
the effective (electrostatic) pair potential by
\begin{equation}
\hat v^{(2)}_{\rm lin}(k)=\hat v_{mm}(k)+\hat v^{(2)}_{\rm ind}(k),
\label{v2link}
\end{equation}
with induced potential
\begin{equation}
\hat v^{(2)}_{\rm ind}(k)=\chi(k)[\hat v_{mc}(k)]^2,
\label{v2indk-cc}
\end{equation}
and the effective triplet potential by
\begin{equation}
\hat v^{(3)}_{\rm eff}({\bf k},{\bf k}')=2\chi'({\bf k}',-{\bf k}-{\bf k}') 
\hat v_{mc}(k)\hat v_{mc}(k')\hat v_{mc}(|{\bf k}+{\bf k}'|),
\label{v3effk-cc}
\end{equation}
where $\chi(k)$ and $\chi'(k)$ are linear and first-order nonlinear 
response functions of the uniform OCP.
Similarly, the first-order corrections for nonlinear response are given by
\begin{equation}
\Delta E=\frac{N_m}{6V_f^2}\left[\sum_{{\bf k},{\bf k}'} \hat
v^{(3)}_{\rm eff}({\bf k},{\bf k}')-N_m\sum_{{\bf k}} \hat
v^{(3)}_{\rm eff}({\bf k},0)\right], 
\label{Delta-E1}
\end{equation}
and
\begin{equation}
\Delta\hat v^{(2)}_{\rm eff}(k)=\frac{1}{V_f}\sum_{{\bf k}'}\hat
v^{(3)}_{\rm eff}({\bf k},{\bf k}')-\frac{N_m}{3V_f}\hat
v^{(3)}_{\rm eff}({\bf k},0),
\label{Delta-v2effk-cc}
\end{equation}
where $V_f=V(1-\eta)$ is the free volume.

\subsubsection{Random Phase Approximation}\label{RPA}

Further progress towards practical expressions for effective interactions
requires specifying the OCP response functions.
For charged colloids, the OCP is typically weakly correlated, 
characterized by relatively small coupling parameters:
$\Gamma=\lambda_B/a_c \ll 1$, where
$\lambda_B=\beta z^2e^2/\epsilon$ is the Bjerrum length and
$a_c=(3/4\pi n_c)^{1/3}$ is the counterion-sphere radius.
For example, for macroions of valence $Z=500$, volume fraction $\eta=0.01$,
and monovalent counterions suspended in salt-free water at room temperature 
($\lambda_B=0.714$ nm), we find $\Gamma\simeq 0.02$.  
For such weakly-correlated plasmas, 
it is reasonable -- at least as regards long-range interactions -- 
to neglect short-range correlations.  We can thus adopt a random phase
approximation (RPA), which equates the two-particle direct
correlation function to its exact asymptotic limit:
$c^{(2)}(r)=-\beta v_{cc}(r)$ or
$\hat c^{(2)}(k)=-4\pi\beta z^2 e^2/\epsilon k^2$. 
Furthermore, we ignore the influence of the macroion hard cores on
the OCP response functions, which is reasonable for
sufficiently dilute suspensions.
Within the RPA, the OCP (two-particle) static structure factor 
and linear response function take the analytical forms
\begin{equation}
S(k)=\frac{1}{1-n_c \hat c^{(2)}(k)}=\frac{1}{1+\kappa^2/k^2}
\label{Sk}
\end{equation}
and
\begin{equation}
\chi(k)=-\beta n_cS(k)=\frac{-\beta n_c}{1+\kappa^2/k^2}, 
\label{chi1}
\end{equation}
where $\kappa=\sqrt{4\pi n_cz^2e^2/\epsilon k_BT}$ is the Debye screening 
constant (inverse screening length), which governs the form of the
counterion density profile and of the screened effective interactions.
In the absence of salt, the counterions are the only screening ions.
The macroions themselves, being singled out as sources of the 
external potential for the counterions, do not contribute to the 
density of screening ions.  
Fourier transforming (\ref{chi1}), the real-space linear response 
function takes the form
\begin{equation}
\chi(r)=-\beta n_c\left[\delta({\bf r})+n_c h_{cc}(r)\right],
\label{chir}
\end{equation}
where
\begin{equation}
h_{cc}(r)=-\frac{\beta z^2e^2}{\epsilon} \frac{e^{-\kappa r}}{r}
\label{hr}
\end{equation}
is the counterion-counterion pair correlation function.
Note the screened-Coulomb (Yukawa) form of $h_{cc}(r)$, with exponential
screening length $\kappa^{-1}$.
Equation~(\ref{chir}) makes clear that there are two physically 
distinct types of counterion response: local response, associated with 
counterion self correlations, and nonlocal response, associated with 
counterion pair correlations.

Proceeding to nonlinear response, we first note that
the three-particle structure factor obeys the identity
\begin{equation}
S^{(3)}({\bf k},{\bf k}')=S(k)S(k')S(|{\bf k}+{\bf k}'|)
\left[1+n_c^2~\hat c^{(3)}({\bf k},{\bf k}')\right],
\label{S3}
\end{equation}
where $\hat c^{(3)}({\bf k},{\bf k}')$ is the Fourier transform of
the three-particle DCF.  Within the RPA, however, $c^{(3)}$ and all 
higher-order DCF's vanish.
Thus, from (\ref{chi1k}), (\ref{chi2k}), and (\ref{S3}), the first 
nonlinear response function can be expressed in Fourier space as
\begin{equation}
\chi'({\bf k},{\bf k}')=-\frac{k_BT}{2n_c^2}\chi(k)\chi(k')
\chi(|{\bf k}+{\bf k}'|)
\label{chi2}
\end{equation}
and in real space as
\begin{equation}
\chi'({\bf r}_1-{\bf r}_2,{\bf r}_1-{\bf r}_3)=-\frac{k_BT}{2n_c^2}
\int{\rm d}{\bf r}\,\chi(|{\bf r}_1-{\bf r}|)\chi(|{\bf r}_2-{\bf r}|)
\chi(|{\bf r}_3-{\bf r}|).
\label{chi2r}
\end{equation}

\subsubsection{Counterion Density Profile}\label{Counterion Density}

An explicit expression for the ensemble-averaged counterion density is 
obtained by substituting the RPA linear response function (\ref{chi1}) 
into the linear response relation 
\begin{equation}
\la\hat\rho_c({\bf k})\ra=\chi(k)\hat v_{\rm ext}(k)
=\chi(k)\hat v_{mc}(k)\hat\rho_m({\bf k}), \qquad k\neq 0.
\label{rhoc1k}
\end{equation}
Inverse transforming (\ref{rhoc1k}) yields
\begin{equation}
\rho_c({\bf r})=\sum_{i=1}^{N_m}\rho_0(|{\bf r}-{\bf R}_i|)
=\int{\rm d}{\bf r}'\,\chi(|{\bf r}-{\bf r}'|)
\sum_{i=1}^{N_m} v_{mc}(|{\bf r}'-{\bf R}_i|),
\label{rhocrlin1}
\end{equation}
which is the real-space linear response counterion density in the presence 
of macroions fixed at positions ${\bf R}_i$, expressed as a sum of 
single-macroion counterion density orbitals $\rho_0(r)$ --
the inverse transform of $\hat\rho_0(k)=\chi(k)\hat v_{mc}(k)$.
Substituting (\ref{chir}) and (\ref{hr}) for the real-space RPA 
linear response function into (\ref{rhocrlin1}), the linear response 
counterion density profile can be expressed as
\begin{equation}
\rho_c({\bf r})=\beta n_c\sum_{i=1}^{N_m}\left[
-v_{mc}(|{\bf r}-{\bf R}_i|)+\frac{\kappa^2}{4\pi} \int {\rm d}{\bf r}'\, 
\frac{e^{-\kappa|{\bf r}-{\bf r}'|}}{|{\bf r}-{\bf r}'|}
v_{mc}(|{\bf r}'-{\bf R}_i|)\right],
\label{rhocr-explicit}
\end{equation}
where the two terms on the right correspond again to local and nonlocal
counterion response.

For hard-core macroions, the form of the macroion-counterion interaction
inside the core is arbitrary and can be specified so as to minimize 
counterion penetration inside the cores~\cite{vRH}.  Thus, assuming
\begin{equation}
v_{mc}(r)=\left\{ \begin{array} {l@{\quad\quad}l}
\frac{\displaystyle -Zze^2}{\displaystyle \epsilon r}, & r>a \\
\frac{\displaystyle -Zze^2}{\displaystyle \epsilon a}\alpha, & r<a
\end{array} \right.
\label{vmcr}
\end{equation}
leaves the freedom to choose the parameter $\alpha$ appropriately.
As shown in \cite{Denton99} and \cite{vRH}, at the level of 
linear response, penetration of counterions inside the macroion cores 
is eliminated by choosing $\alpha=\kappa a/(1+\kappa a)$.  
This choice yields
\begin{equation}
\hat v_{mc}(k)=-\frac{4\pi Zze^2}{\epsilon(1+\kappa a)k^2}
\left[\cos(ka)+\frac{\kappa}{k} \sin(ka) \right]
\label{vmck2}
\end{equation}
and
\begin{equation}
\rho_0(r)=\left\{ \begin{array}
{l@{\quad\quad}l}
\frac{\displaystyle Z}{\displaystyle z}\frac{\displaystyle \kappa^2}
{\displaystyle 4\pi}
~\frac{\displaystyle e^{\kappa a}}{\displaystyle 1+\kappa a}
~\frac{\displaystyle e^{-\kappa r}}{\displaystyle r}, & r>a \\
0, & r<a,
\end{array} \right.
\label{rho1r2}
\end{equation}
which agrees precisely with the asymptotic ($r\to\infty$) expression 
predicted by the DLVO theory of charged colloids~\cite{DL,VO}.

\subsubsection{Effective Electrostatic Interactions}\label{EEI}

Practical expressions for the effective electrostatic interactions 
are obtained by explicitly evaluating inverse Fourier transforms.
By combining (\ref{Elin-cc}), (\ref{v2indk-cc}), (\ref{v3effk-cc}), 
(\ref{Delta-E1}), and (\ref{vmck2}), the volume energy can be expressed 
as the sum of the linear response approximation,
\begin{equation}
E_{\rm lin}=F_{\rm OCP}-N_m\frac{Z^2e^2}{2\epsilon}
\frac{\kappa}{1+\kappa a}-\frac{N_ck_BT}{2}, 
\label{E02}
\end{equation}
and the first nonlinear correction,
\begin{equation}
\Delta E=-\frac{N_mk_BT}{6n_c^2}\left(\int{\rm d}{\bf r}\,
\left[\rho_0(r)\right]^3
-n_c\int{\rm d}{\bf r}\,\left[\rho_0(r)\right]^2\right).
\label{Delta-E1r}
\end{equation}
The first term on the right side of (\ref{E02}) accounts for the 
counterion entropy and the second term for the macroion-counterion 
electrostatic interaction energy.  The latter term happens to be identical
to the energy that would result if each macroion's counterions were 
all concentrated at a distance of one screening length ($\kappa^{-1}$) 
away from the macroion surface.  

From (\ref{v2link}), (\ref{v2indk-cc}), (\ref{chi1}) and (\ref{vmck2}), 
the linear response prediction for the effective pair interaction is given by
\begin{equation}
v_{\rm lin}^{(2)}(r)=\frac{Z^2e^2}{\epsilon}\left(\frac{e^{\kappa a}}
{1+\kappa a}\right)^2~\frac{e^{-\kappa r}}{r}, \qquad r>\sigma, 
\label{v2linr}
\end{equation}
which is identical to the familiar DLVO screened-Coulomb potential 
in the dilute limit of widely separated macroions~\cite{DL,VO}, 
while (\ref{Delta-v2effk-cc}) yields the first nonlinear correction 
\begin{equation}
\Delta v^{(2)}_{\rm eff}(r)=-\frac{k_BT}{n_c^2}\int{\rm d}{\bf r}'\,
\rho_0(r')\rho_0(|{\bf r}-{\bf r}'|)\left[\rho_0(|{\bf r}-{\bf r}'|)
-\frac{n_c}{3}\right]. 
\label{Delta-v2effr-2}
\end{equation}
Finally, from (\ref{v3effk-cc}) and (\ref{chi2}), the effective triplet 
interaction is
\begin{equation}
v^{(3)}_{\rm eff}({\bf r}_{12},{\bf r}_{13})
=-\frac{k_BT}{n_c^2}\int{\rm d}{\bf r}\,\rho_0(|{\bf r}_1-{\bf r}|)
\rho_0(|{\bf r}_2-{\bf r}|)\rho_0(|{\bf r}_3-{\bf r}|).
\label{v3effr-2}
\end{equation}
Note that the final terms on the right sides of 
(\ref{E02}), (\ref{Delta-E1r}), and (\ref{Delta-v2effr-2})
originate from the charge neutrality constraint.

The above results generalize straightforwardly to nonzero salt concentration.
Here we merely sketch the steps leading to the final expressions, referring 
the reader to \cite{Denton00} and \cite{Denton-pre04} for details.
Assuming fixed average number density (in the free volume) of salt ion pairs,
$n_s=N_s/V_f$, the total average microion density is $n_{\mu}=n_++n_-=n_c+2n_s$,
where $n_{\pm}$ are the average number densities of positive/negative microions.
Following \cite{Denton00}, the Hamiltonian generalizes to
$H=H_{mm}+H_{\mu}+H_{m+}+H_{m-}$,
where $H_{\mu}$ is the Hamiltonian of all microions (counterions and salt ions) 
and $H_{m\pm}$ are the electrostatic interaction energies between macroions 
and positive/negative microions.  The perturbation theory proceeds as before, 
except that the reference system is now a two-component plasma.
The presence of positive and negative microion species entails a
proliferation of response functions, $\chi_{ij}$ and $\chi'_{ijk}$, 
$i,j,k=\pm$, and a generalization of (\ref{chir}) to
\begin{eqnarray}
\chi_{++}(r)&=&-\beta n_+\left[\delta({\bf r})+n_+ h_{++}(r)\right]
\label{chi++r}
\nonumber \\
\chi_{+-}(r)&=&-\beta n_+n_- h_{+-}(r)
\label{chi+-r}
\\
\chi_{--}(r)&=&-\beta n_-\left[\delta({\bf r})+n_- h_{--}(r)\right],
\label{chi--r}
\nonumber 
\end{eqnarray}
where $h_{ij}(r)$, $i,j=\pm$, are the two-particle pair correlation 
functions of the microion plasma.
Substituting the ensemble-averaged microion number densities into the 
multi-component Hamiltonian yields expressions for the macroion-microion 
interaction terms and, in turn, the effective interactions.

The ultimate effect of salt is to modify the previous results as follows.
First, the average counterion density that appears in the Debye screening 
constant and in the linear response function (\ref{chi1}) is replaced 
by the total average microion density: 
$\kappa=\sqrt{4\pi n_{\mu}z^2e^2/\epsilon k_BT}$ and
$\chi(k)=-\beta n_{\mu}S(k)$.  The first nonlinear response function retains 
its original form (\ref{chi2}), but with the new definition of $\kappa$
and with $n_c$ replaced by $n_{\mu}$.  Second, the linear response 
volume energy becomes~\cite{Denton00}
\begin{equation}
E_{\rm lin}=F_{\rm plasma}-N_m\frac{Z^2e^2}{2\epsilon}
\frac{\kappa}{1+\kappa a}-\frac{(N_+-N_-)^2}{N_++N_-}\frac{k_BT}{2},
\label{E03}
\end{equation}
where 
\begin{equation}
F_{\rm plasma}=k_BT\{N_+[\ln(n_+\Lambda_+^3)-1]+N_-[\ln(n_-\Lambda_-^3)-1]\}
\label{Fplasma}
\end{equation}
is the free energy of the unperturbed microion plasma (in the free volume) and
$\Lambda_{\pm}$ denotes the thermal wavelengths of positive and negative 
microions.  Third, the effective triplet interaction and nonlinear corrections 
to the effective pair interaction and volume energy are generalized as follows:
\begin{equation}
\beta\Delta E=-\frac{N_m}{6}\frac{(n_+-n_-)}{n_{\mu}^3}\left(
\int{\rm d}{\bf r}\,\left[\rho_0(r)\right]^3
-n_{\mu}\int{\rm d}{\bf r}\,\left[\rho_0(r)\right]^2\right)
\label{Delta-E1r-salt}
\end{equation}
\begin{equation}
\beta\Delta v^{(2)}_{\rm eff}(r)=-\frac{(n_+-n_-)}{n_{\mu}^3}\int{\rm d}
{\bf r}'\,\rho_0(r')\rho_0(|{\bf r}-{\bf r}'|)\left[\rho_0(|{\bf r}
-{\bf r}'|)-\frac{n_{\mu}}{3}\right]
\label{Delta-v2effr-salt}
\end{equation}
\begin{equation}
\beta v^{(3)}_{\rm eff}({\bf r}_{12},{\bf r}_{13})=-\frac{(n_+-n_-)}{n_{\mu}^3}
\int{\rm d}{\bf r}\, \rho_0(|{\bf r}_1-{\bf r}|)\rho_0(|{\bf r}_2-{\bf r}|)
\rho_0(|{\bf r}_3-{\bf r}|). 
\label{v3effr-salt}
\end{equation}
These results imply that nonlinear effects increase in strength with 
increasing charge and concentration of macroions and with 
decreasing salt concentration, and that effective triplet interactions 
are consistently attractive.  
It is also clear that in the limit of zero macroion concentration 
($n_c=n_+-n_-\to 0$), or of high salt concentration ($n_{\mu}\to\infty$), 
such that $(n_+-n_-)/n_{\mu}\to 0$, the leading-order nonlinear corrections 
all vanish.  This result -- a consequence of charge neutrality -- may 
partially explain the remarkably broad range of validity of DLVO theory 
for suspensions at high ionic strength.

The wide tunability of the effective electrostatic interactions leads to 
rich phase behaviour in charge-stabilised colloidal suspensions.
Simulation studies~\cite{Stevens96} of one-component systems interacting via
the screened-Coulomb pair potential have demonstrated that variation in the 
Debye screening constant -- corresponding in experiments to varying salt 
concentration -- can account for the observed cross-over in relative stability 
between stable fcc and bcc crystal structures. 
The density dependence of the volume energy, resulting from the constraints 
of fixed density and charge neutrality, can have profound implications for 
thermodynamic properties (e.g., phase behaviour, osmotic pressure) of highly
deionized suspensions.  Specifically, the volume energy has been predicted
by DFT~\cite{vRH,vRDH}, extended Debye-H\"uckel theory~\cite{Warren}, and
response theory~\cite{Denton-pre06} to drive an unusual counterion-induced 
phase separation between macroion-rich and macroion-poor phases at low 
(sub-mM) salt concentrations.  Despite purely repulsive pair interactions, 
which oppose bulk phase separation, counterion entropy and macroion self-energy
can, according to predictions, conspire to drive a spinodal instability.  
It remains unresolved, however, whether this predicted instability is related 
to experimental observations of anomalous phase behaviour in charged colloids, 
including stable voids~\cite{Tata} and metastable crystallites~\cite{Grier97}.

Predictions of response theory can be directly tested against simulations. 
As an example, Fig.~\ref{fig-effint} presents a comparison~\cite{Denton-pre04} 
with available data from {\it ab initio} simulations~\cite{Tehver} for 
the total potential energy of interaction between a pair of macroions, 
of diameter $\sigma=106$ nm and valence $Z=200$, in a cubic box of 
length 530 nm (with periodic boundary conditions) in the absence of salt.  
The theory is in excellent agreement with simulation, although nonlinear 
effects are relatively weak for these parameters.  
Figure~\ref{fig-effint} also illustrates the effective triplet interaction 
between a trio of macroions arranged in an equilateral triangle for 
$\sigma=100$ nm and two different valences, $Z=500$ and 700, computed 
from ({\ref{v3effr-salt})~\cite{Denton-pre04}.  The strength of the 
attractive interaction grows rapidly with increasing macroion valence and 
with decreasing separation between macroion cores.  Other methods, including 
DFT~\cite{Lowen-Allahyarov} and Poisson-Boltzmann theory~\cite{Russ02},
predict qualitatively similar triplet interactions.
In concentrated suspensions of highly-charged macroions, higher-order
effective interactions may become significant.  
\begin{figure}
\centering
\begin{minipage}{5cm}
\includegraphics[height=4.2cm]{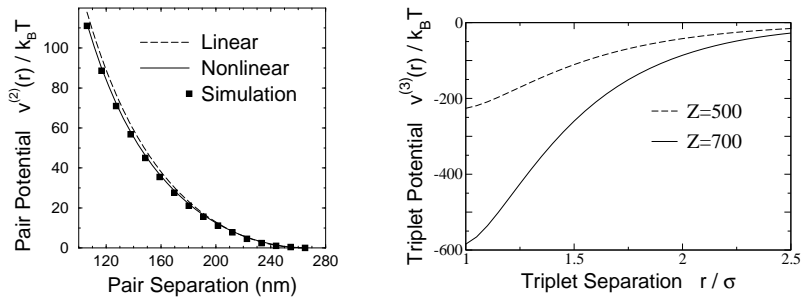}
\end{minipage}
\begin{minipage}{5cm}
\includegraphics[height=3.8cm]{v3r.eps}
\end{minipage}
\caption{\label{fig-effint} 
Left: Interaction energy of two macroions (diameter $106$ nm, valence $200$) 
in a cubic box of length 530 nm (with periodic boundary conditions) at zero 
salt concentration.
The potentials are shifted to zero at maximum macroion separation.
Dashed curve: linear response prediction. 
Solid curve: nonlinear response prediction~\cite{Denton-pre04}.
Symbols: {\it ab initio} simulation data~\cite{Tehver}.
Right: Effective triplet interaction between three macroions, 
arranged in an equilateral triangle of side length $r$, 
with macroion diameter $\sigma=100$ nm, macroion valence $Z=500$ 
(dashed curve) or $Z=700$ (solid curve), volume fraction $\eta=0.01$, 
and salt concentration $c_s=1$ $\mu$M.
Computed from (\ref{v3effr-salt}) \cite{Denton-pre04}.}
\end{figure}

\subsection{Systems of Hard Particles}\label{Hard}

Soft matter systems that contain hard (impenetrable) particles include 
colloidal and nanoparticle dispersions and colloid-nanoparticle mixtures.  
In such systems, the bare interparticle interactions depend, at least in part, 
on the geometric volume excluded by each particle to all other particles.  
In general, van der Waals, electrostatic, and other interactions may also 
be present.  The simplest case is that of spherical particles, as in 
bidisperse or polydisperse mixtures of colloids and/or nanoparticles.  In a 
binary ($ab$) hard-sphere mixture, the bare pair interactions have the form
\begin{equation}
v_{\alpha\beta}(r)=\left\{ \begin{array} {l@{\quad\quad}l}
\infty, & r<R_{\alpha}+R_{\beta} \\
0, & r\ge R_{\alpha}+R_{\beta},
\end{array} \right. 
\label{vhs}
\end{equation}
where $\alpha,\beta=a,b$ and $R_{\alpha}$ denotes the radius of particles 
of type $\alpha$.  For other shapes, the pair interactions naturally
depend also on the orientations of the particles.
The methods outlined in Sec.~\ref{Interactions} for modelling effective 
interactions are easily adapted to hard-particle systems.  In what follows, 
we first present general formulae and then describe an application to a 
common model of colloid-polymer mixtures.

\subsubsection{Response Theory for Excluded-Volume Interactions}
\label{Response-EVI}

As noted in Sec.~\ref{Perturbation}, the perturbative response theory is 
most useful in cases where the mesoscopic particles possess a property that 
can be continuously varied to turn on the ``external" potential for the
other (microscopic) particles.  In the case of electrostatic interactions,
the obvious tunable property is the charge on the particles.  Analogously, 
for excluded-volume interactions, the relevant property is the volume
occupied by the particles.  In an $ab$ mixture, we can imagine the mesoscopic 
($a$) particles to be inflated continuously from points to their full size.
To grow in size, the $a$ particles must push against the surrounding $b$ 
particles.  As they grow, the $a$ particles sweep out spheres, of radius
$R_a+R_b$, from which the centres of the $b$ particles are excluded.
The change in free energy during this process equals the 
reversible work performed by the expanding $a$ particles against the 
osmotic pressure exerted by the $b$ fluid.  Adapting (\ref{Fb2}) to 
quantify this conceptual image, the Helmholtz free energy of the $b$ particles 
in the presence of the $a$ particles can be expressed as
\begin{equation}
F_b=F_0+\int_0^1{\rm d}\lambda\, 
V_{\rm exc}[\rho_a;\lambda]~\Pi_b[\rho_a,\rho_b;\lambda],
\label{Fb-hard1}
\end{equation}
where $F_0$ is again the free energy of the unperturbed (reference) $b$ fluid, 
the $\lambda$ integral continuously scales the $a$ particles from points to 
full size, and $V_{\rm exc}[\rho_a;\lambda]$ and $\Pi_b[\rho_a,\rho_b;\lambda]$ 
-- both functionals of density -- are, respectively, the total volume excluded 
to the $b$ particles, and the osmotic pressure exerted by the $b$ fluid, in the 
presence of $a$ particles expanded to a fraction $\lambda$ of their full volume.

Although (\ref{Fb-hard1}) is formally exact for any shape of particle, the 
complicated dependence of the excluded volume and osmotic pressure on the scale 
parameter $\lambda$ precludes an exact evaluation of $F_b$ for an arbitrary 
configuration of $a$ particles.  One approximation scheme is based on 
expanding the osmotic pressure in powers of $\lambda$.  
Expanding around $\lambda=0$, for example, yields
\begin{equation}
F_b=F_0+\Pi_b^{(0)}(n_b)V_{\rm exc}[\rho_a]+
\left(\frac{\partial\Pi_b}{\partial\lambda}\right)_0
\int_0^1{\rm d}\lambda\, \lambda V_{\rm exc}[\rho_a;\lambda]+\cdots,
\label{Fb-hard2}
\end{equation}
where $\Pi_b^{(0)}(n_b)$ is the osmotic pressure of the reference fluid and 
$V_{\rm exc}[\rho_a]$ is the total volume excluded to the $b$ particles by 
full-sized $a$ particles.  In practical applications, such as colloid-polymer 
mixtures, the system is often coupled to an infinite reservoir of $b$ 
particles (e.g., a polymer solution) that fixes the chemical potential $\mu_b$ 
of the $b$ component.  In this case, a more natural choice for $\Pi_b^{(0)}$ 
may be the osmotic pressure of the reservoir $\Pi_b^{(r)}$, leading to the 
approximation
\begin{equation}
F_b=F_0+\Pi_b^{(r)}V_{\rm exc}[\rho_a].
\label{Fb-hard3}
\end{equation}
The higher-order terms in (\ref{Fb-hard2}), which depend on $bb$ pair
interactions, are difficult to evaluate and commonly ignored.  
Nevertheless, (\ref{Fb-hard3}) is often a reasonable approximation, 
especially in the dilute limit ($n_a \to 0$), where 
$\Pi_b[\rho_a,\rho_b;\lambda]$ depends weakly on $n_a$ and $\lambda$.  
In the idealized case in which the $b$ fluid can be modelled as a 
noninteracting gas, the osmotic pressures of the system and reservoir 
are equal and (\ref{Fb-hard3}) then becomes exact (see below).

It remains a highly nontrivial problem, for an arbitrary configuration 
of $a$ particles, to approximate the excluded volume $V_{\rm exc}[\rho_a]$, 
which requires determining the intersection volumes of the mutually overlapping
exclusion spheres that surround the $a$ particles.  In the simplest case of 
non-overlapping spheres, the total excluded volume is merely the excluded 
volume of a single $a$ particle times the number of particles.  In the next 
simplest case, in which only pairs of exclusion spheres overlap, the 
intersection volumes of overlapping pairs must be subtracted to avoid 
double-counting.  The next step corrects for the intersection volume of 
three mutually overlapping spheres.  The geometrical problem is illustrated 
in Fig.~\ref{fig-depletion} for a mixture of hard spherical colloids and 
coarse-grained spherical polymers.  Expressing the total excluded volume as 
a sum of overlap terms, and assuming isotropic interactions (spherical 
particles), the free energy can be approximated by
\begin{equation}
F_b=F_0+\Pi_b^{(r)}\left(N_a V_{\rm exc}^{(1)}-\frac{1}{2}\sum_{i\neq j=1}^{N_a}
V_{\rm ov}^{(2)}(r_{ij})+\frac{1}{3!}\sum_{i\neq j\neq k=1}^{N_a}
V_{\rm ov}^{(3)}({\bf r}_{ij},{\bf r}_{ik})-\cdots\right),
\label{Fb-hard4}
\end{equation}
where $V_{\rm exc}^{(1)}$ is the excluded volume of a single $a$ particle,
$V_{\rm ov}^{(2)}(r_{ij})$ is the intersection volume of a pair of 
overlapping exclusion spheres surrounding particles $i$ and $j$, and
$V_{\rm ov}^{(3)}({\bf r}_{ij},{\bf r}_{ik})$ is the intersection volume 
of three mutually overlapping spheres surrounding particles $i$, $j$, and $k$.
In the more general case of anisotropic interactions (nonspherical particles), 
the effective interactions depend also on the relative orientations
of the particles.
\begin{figure}
\centering
\begin{minipage}{5cm}
\includegraphics[height=4.2cm]{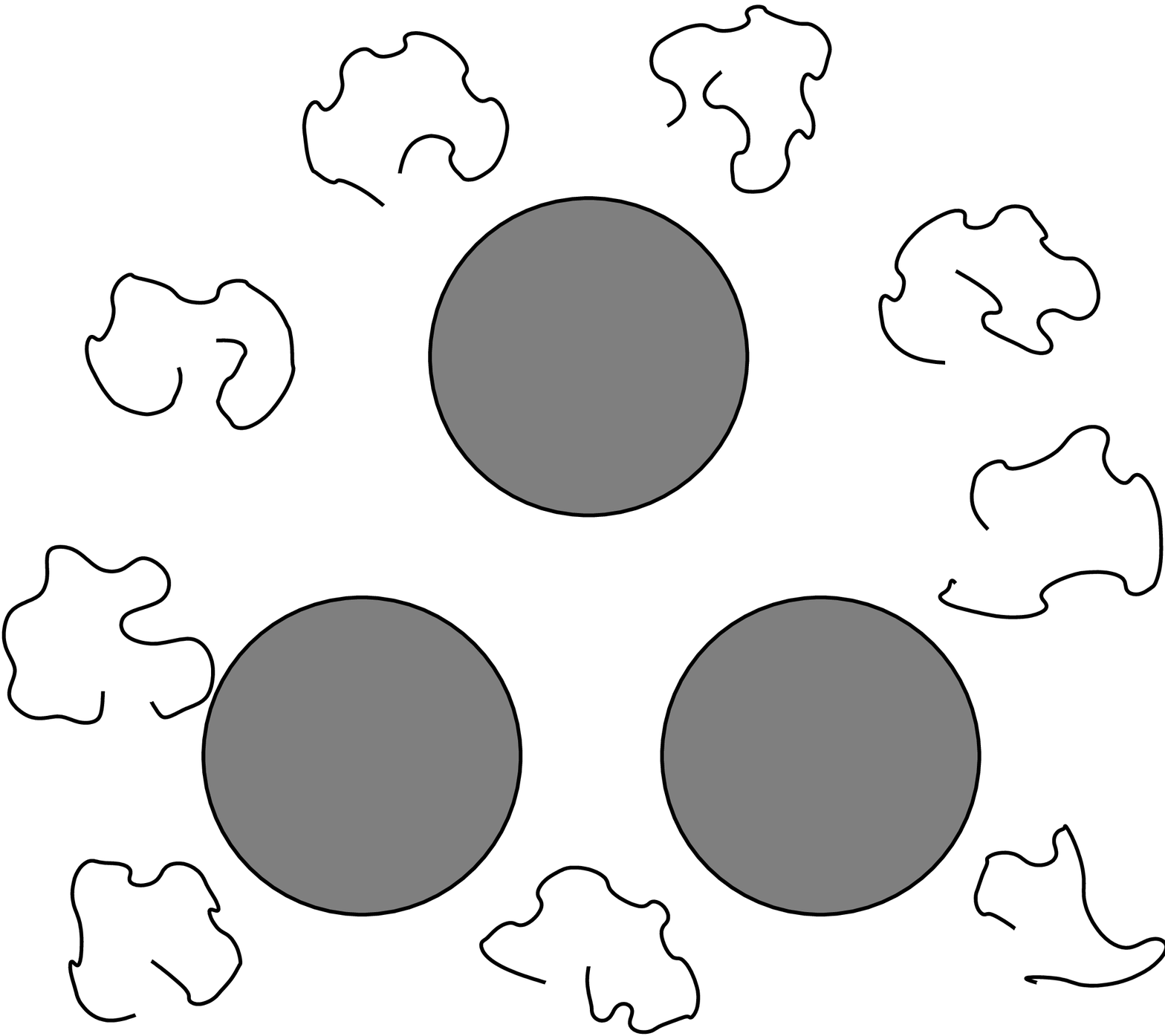}
\end{minipage}
\hspace*{0.5cm}
\begin{minipage}{5cm}
\includegraphics[height=3.8cm]{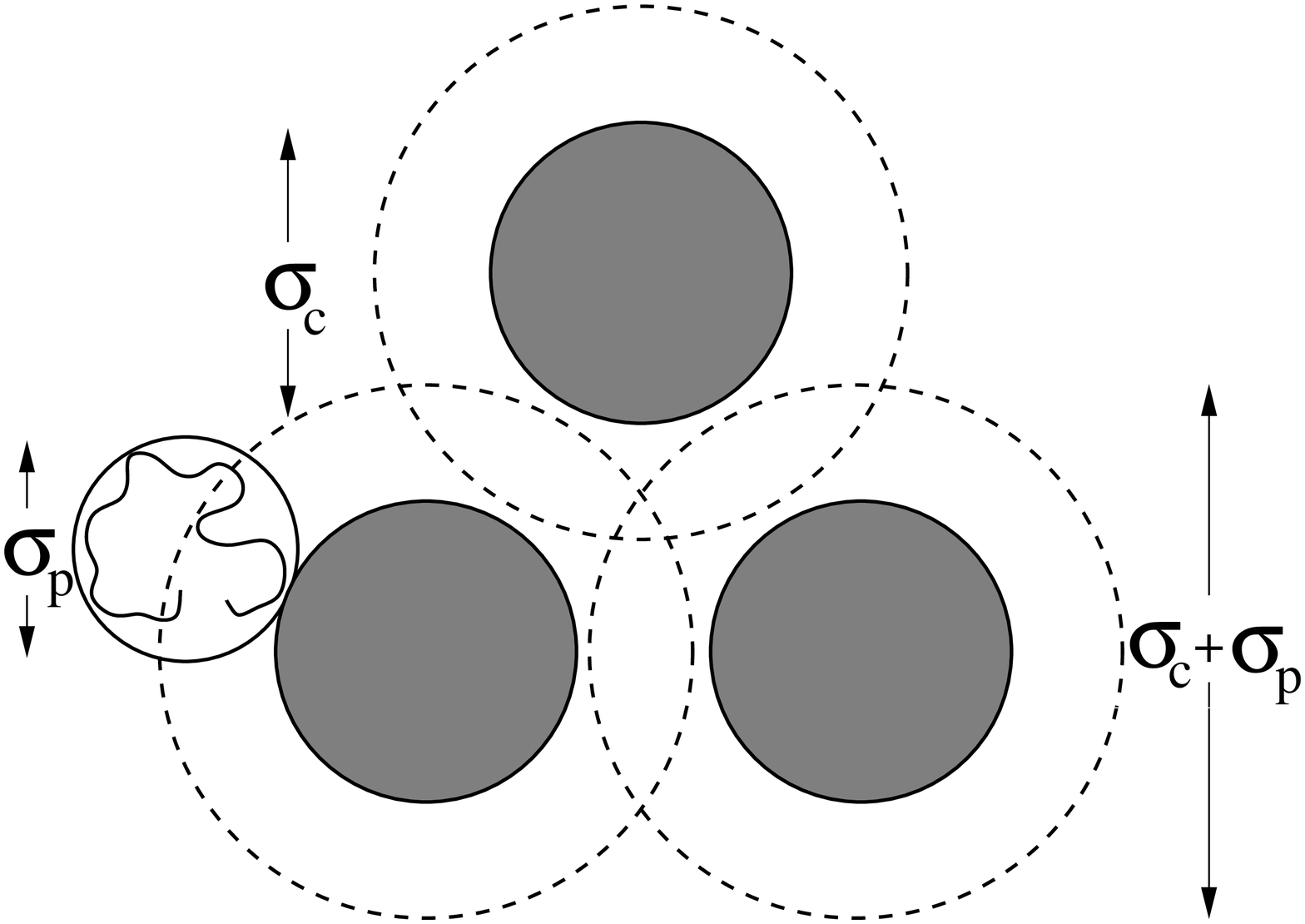}
\end{minipage}
\caption{\label{fig-depletion} 
Left: Colloid-polymer mixture with hard (excluded-volume) interactions.  Depletion 
of polymer coils from spaces between colloids induces effective interactions 
between colloids.  Right: Asakura-Oosawa-Vrij model with colloids treated as 
hard spheres (diameter $\sigma_c$) and polymers as coarse-grained spheres 
(diameter $\sigma_p$).  The effective interactions are a function of the 
total volume excluded by the colloids to the polymer centres, 
which depends on the intersection volumes of mutually overlapping spheres 
of exclusion (diameter $\sigma_c+\sigma_p$) surrounding each colloid.}
\end{figure}

From (\ref{Fb-hard4}), the effective Hamiltonian is seen to have the
same general form as in (\ref{Heff2}), with a one-body volume term
\begin{equation}
E=F_0+\Pi_b^{(r)} N_a V_{\rm exc}^{(1)},
\label{E-hard}
\end{equation}
an effective pair potential
\begin{equation}
v_{\rm eff}^{(2)}(r)=v_{aa}(r)-\Pi_b^{(r)} V_{\rm ov}^{(2)}(r),
\label{v2eff-hard}
\end{equation}
and an effective triplet potential
\begin{equation}
v_{\rm eff}^{(3)}({\bf r}_{ij},{\bf r}_{ik})=
\Pi_b^{(r)} V_{\rm ov}^{(3)}({\bf r}_{ij},{\bf r}_{ik}).
\label{v3eff-hard}
\end{equation}
This simple low-density approximation exhibits several noteworthy features.
The induced pair potential has the same sign as its electrostatic counterpart 
(negative and attractive), while the effective triplet potential has the 
opposite sign (positive and repulsive).  The induced pair attraction originates
from the depletion of $b$ particles from the space between pairs of closely 
approaching $a$ particles and the resulting imbalance in osmotic pressure.
Furthermore, in contrast to systems of charged particles with electrostatic 
interactions, in systems of hard particles with excluded-volume interactions,
the volume term depends only trivially on the density of the $a$ particles
and the effective many-body interactions do not generate corrections to 
lower-order effective interactions.  One must take care, however, not to draw 
general conclusions, since the neglected higher-order terms in 
(\ref{Fb-hard2}) can significantly modify the effective interactions -- 
especially in concentrated systems -- introducing density-dependence and even 
changing the sign. 
In binary hard-sphere mixtures, for example, packing of smaller spheres around 
larger spheres induces effective pair interactions between the larger spheres 
that can exhibit a repulsive barrier and even long-range 
oscillations~\cite{Dijkstra-pre99,Goetzelmann99}.

A simple, but important, example of a binary mixture of spheres is the 
Asakura-Oosawa-Vrij (AOV) model~\cite{AO,Vrij} of mixtures of colloids 
and free (nonadsorbing) polymers.  The AOV model treats the colloids 
($a\leftrightarrow c$) as hard spheres, interacting via an additive 
hard-sphere pair potential $v_{cc}(r)$ (\ref{vhs}), and the polymers 
($b\leftrightarrow p$) as effective, coarse-grained spheres that have 
hard interactions with the colloids, 
\begin{equation}
v_{cp}(r)=\left\{ \begin{array} {l@{\quad\quad}l}
\infty, & r<R_c+R_p \\
0, & r\ge R_c+R_p,
\end{array} \right. 
\label{vcc}
\end{equation}
but are mutually noninteracting (ideal): $v_{pp}(r)=0$ for all $r$.  The
radius $R_p$ of the effective polymer spheres is most naturally identified 
with the polymer radius of gyration.  The neglect of polymer-polymer
interactions is strictly valid only for theta solvents~\cite{deGennes79}, 
wherein monomer-monomer excluded-volume interactions effectively vanish.
In this special case, the polymer in the system behaves as an ideal gas
confined to the {\it free} volume, $V_f=\alpha V$, where the free volume 
fraction $\alpha$ is defined as the ratio of the volume available to the 
polymer centres (i.e., not excluded by the hard colloids) to the total volume.
At equilibrium, equality of the polymer chemical potentials in the system, 
$\mu_p=k_BT\ln(n_p\Lambda_p^3/\alpha)$, and in the reservoir, 
$\mu_p^{(r)}=k_BT\ln(n_p^{(r)}\Lambda_p^3)$,
implies that the corresponding polymer densities must be related via
$n_p=\alpha n_p^{(r)}$.  This simple relation imposes equality also 
of the polymer osmotic pressures in the system and reservoir, 
$\Pi_p=\Pi_p^{(r)}=k_BTn_p^{(r)}$, 
thus rendering (\ref{Fb-hard3}) and (\ref{v2eff-hard}) exact.  

Within the AOV model, not only are the effective interactions exact, 
but the effective one-body and pair interactions have analytical forms, 
since the single-sphere excluded volume is simply 
$V_{\rm exc}^{(1)}=(4\pi/3)(R_a+R_b)^3$, and the convex-lens-shaped 
pair intersection has a volume
\begin{equation}
V_{\rm ov}^{(2)}(r)=\left\{ \begin{array} {l@{\quad}l}
\frac{\displaystyle \pi}{\displaystyle 6}
\left(\sigma_p+\sigma_c\right)^3
\left[1-\frac{\displaystyle 3r}{\displaystyle 2(\sigma_c+\sigma_p)}+
\frac{\displaystyle r^3}{\displaystyle 2(\sigma_c+\sigma_p)^3}\right], 
& \sigma_c<r<\sigma_c+\sigma_p \\
0, & r\geq\sigma_c+\sigma_p,
\end{array} \right.
\label{v2overlap}
\end{equation}
where $\sigma_c=2R_c$ and $\sigma_p=2R_p$ are the particle diameters and
$q=R_p/R_c$ is the size ratio.  The effective one-body interaction
(\ref{E-hard}), being linear in the colloid density, does not affect 
phase behaviour, but does contribute to the total osmotic pressure.  
The effective pair potential described by
(\ref{v2eff-hard}) and (\ref{v2overlap}) consists of a repulsive 
hard-sphere core and an attractive well with a range equal to the sum of the 
particle diameters and depth proportional to the reservoir polymer osmotic 
pressure.  For sufficiently small size ratios ($q\le 0.154$), such that 
the spherical exclusion spheres surrounding three colloids never intersect, 
the effective triplet (and higher-order) interactions are identically zero.  
For $q>0.154$, the effective triplet interaction is nonzero and repulsive.  

Although limited to ideal polymers, the AOV model gives qualitative insight 
even for interacting polymers.  For sufficiently large size ratios ($q>0.45$) 
and high polymer concentrations, polymer-depletion-induced effective pair 
attractions can drive demixing into colloid-rich (polymer-poor) and 
colloid-poor (polymer-rich) fluid phases~\cite{Pusey}.  Simulations of the
effective Hamiltonian system~\cite{Dijkstra-jpcm99} and of the full binary
AO model~\cite{Meijer-Frenkel} indicate that for smaller size ratios 
($q\le 0.45$), fluid-fluid demixing is only metastable, being preempted 
by the fluid-solid (freezing) transition.  With the phase diagram of the AOV
model now well understood, recent attention has turned to exploring the 
phase behaviour of mixtures of colloids and interacting polymers 
(Sec.~\ref{Soft}).

\subsubsection{Cluster Expansion Approach}\label{Cluster}

An alternative, and elegant, approach to modelling effective interactions 
in systems of hard particles has been developed recently by Dijkstra 
{\it et al}~\cite{Dijkstra-jpcm99,Dijkstra-pre99}.  This powerful statistical 
mechanical method is similarly based on integrating out the degrees of 
freedom of one species of particle in the presence of fixed particles of 
another species.  The essence of the method is perhaps most transparent in the 
context of the AOV model of colloid-polymer mixtures~\cite{Dijkstra-jpcm99}, 
defined by the Hamiltonian 
$H=K+H_{cc}+H_{cp}$, where $K$ is the kinetic energy and
\begin{equation}
H_{cc}=\frac{1}{2}\sum_{{i\neq j=1}}^{N_c}v_{cc}(r_{ij}), \qquad
H_{cp}=\sum_{{i=1}}^{N_c}\sum_{{j=1}}^{N_p}v_{cp}(r_{ij})
\label{Hcc}
\end{equation}
describe the colloid-colloid and colloid-polymer interactions.  Assuming an 
infinite reservoir that exchanges polymer with the system, the natural choice 
of ensemble is the semigrand canonical ensemble, in which the number of 
colloids $N_c$, volume $V$, temperature $T$, and polymer chemical potential 
$\mu_p$ are fixed.  Within this ensemble, which treats the colloids canonically 
and the polymers grand canonically, the appropriate thermodynamic potential 
is the semigrand potential $\Omega$, given by:
\begin{eqnarray}
\exp(-\beta\Omega)&=&\langle\langle\exp[-\beta(H_{cc}+H_{cp})]
\rangle_p\rangle_c 
\nonumber \\
&=&\frac{1}{N_c!\Lambda_c^{3N_c}}\int{\rm d}{\bf r}^{N_c}\, \exp(-\beta H_{cc})
\sum_{N_p=0}^{\infty}\frac{z_p^{N_p}}{N_p!}
\int{\rm d}{\bf r}^{N_p}\, \exp(-\beta H_{cp})
\nonumber \\
&=& \la\exp(-\beta H_{\rm eff})\ra_c, 
\label{omega}
\end{eqnarray}
where $\la\cdots\ra_c$ and $\la\cdots\ra_p$ denote semigrand canonical traces 
over colloid and polymer coordinates, $z_p=e^{\beta\mu_p}/\Lambda_p^3$ is the 
polymer fugacity, and $\Lambda_c$ and $\Lambda_p$ are the colloid and polymer 
thermal wavelengths.  Here $H_{\rm eff}=H_{cc}+\Omega_p$ is the 
effective one-component Hamiltonian, where $\Omega_p$ represents the grand 
potential of the polymers in the presence of fixed colloids, defined by
\begin{eqnarray}
\exp(-\beta\Omega_p)&=&\la\exp(-\beta H_{cp})\ra_p \nonumber \\
&=&\sum_{N_p=0}^{\infty}\frac{z_p^{N_p}}{N_p!}\left(\int{\rm d}{\bf r}\, 
\exp\left[-\beta\sum_{{i=1}}^{N_c}v_{cp}(|{\bf r}_i-{\bf r}|)\right]
\right)^{N_p},
\nonumber \\
&=&\exp\left(z_p\int{\rm d}{\bf r}\, 
\exp\left[-\beta\sum_{{i=1}}^{N_c}v_{cp}(|{\bf r}_i-{\bf r}|)\right]\right).
\label{omegap1}
\end{eqnarray}
Equating arguments of the exponential functions on the left and right sides,
we have
\begin{equation}
-\beta\Omega_p=z_p\int{\rm d}{\bf r}\, \exp\left[-\beta\sum_{{i=1}}^{N_c}
v_{cp}(|{\bf r}_i-{\bf r}|)\right].
\label{omegap2}
\end{equation}

As shown by Dijkstra {\it et al}~\cite{Dijkstra-jpcm99,Dijkstra-pre99}, the 
polymer grand potential can be systematically approximated by a cluster
expansion technique drawn from the theory of simple liquids~\cite{HM}. 
Defining the Mayer functions
\begin{equation}
f_{ij}\equiv\exp[-\beta v_{cp}(r_{ij})]-1=
\left\{ \begin{array} {l@{\quad\quad}l}
-1, & r_{ij}<R_c+R_p \\
~~0, & r_{ij}\ge R_c+R_p,
\end{array} \right. 
\label{mayer}
\end{equation}
(\ref{omegap2}) can be expanded as follows:
\begin{eqnarray}
-\beta\Omega_p&=&z_p\int{\rm d}{\bf r}_j\, \prod_{i=1}^{N_c}(1+f_{ij}) 
\nonumber \\
&=&z_p\int{\rm d}{\bf r}_j\, \left(1+\sum_{{i=1}}^{N_c}f_{ij} 
+\sum_{{i<k=1}}^{N_c}f_{ij}f_{kj}+\cdots\right). 
\label{omegap3}
\end{eqnarray}
Equation~(\ref{omegap3}) is a form of cluster expansion well-suited to 
systematic approximation by diagrammatic techniques~\cite{HM}.  Successive 
terms in the summation, generated by increasing numbers of colloids 
interacting with the polymers, correspond directly to the one-body volume 
term and effective pair and many-body interactions.  
In fact, from (\ref{omegap3}), the semigrand potential can be written 
in a form that is precisely analogous to the free energy defined in 
(\ref{Fb-hard3}):
\begin{equation}
\beta\Omega_p=\beta\Omega_0+z_p\left(N_c V_{\rm exc}^{(1)}
-\frac{1}{2}\sum_{i\neq j=1}^{N_c}V_{\rm ov}^{(2)}(r_{ij})
+\frac{1}{3!}\sum_{i\neq j\neq k=1}^{N_c}
V_{\rm ov}^{(3)}({\bf r}_{ij},{\bf r}_{ik})-\cdots\right),
\label{omegap4}
\end{equation}
where $\beta\Omega_0=-z_pV$ is the grand potential of the reference system
of pure polymer.  In the case of ideal polymer, the polymer fugacity is 
simply related to the reservoir osmotic pressure via $z_p=\beta\Pi_p^{(r)}$. 
Note that (\ref{Fb-hard3}) and (\ref{omegap4}) are completely equivalent, 
differing only with respect to the relevant ensemble, with
(\ref{Fb-hard3}) applying in the canonical ensemble and (\ref{omegap4}) 
in the semigrand canonical ensemble.  

For mixtures of colloids and polydisperse polymers, (\ref{omegap3}) 
generalizes to
\begin{equation}
-\beta\Omega_p=\sum_k z_p^{(k)}\int{\rm d}{\bf r}_j\, \prod_{i=1}^{N_c}
(1+f_{ij}^{(k)}), 
\label{omegap5}
\end{equation}
where $z_p^{(k)}$ is the fugacity of polymer species $k$ and $f_{ij}^{(k)}$
is the corresponding Mayer function.  From (\ref{omegap5}), the effective
pair potential is then simply a sum of depletion potentials, each induced by 
a polymer species of a different size.  On the other hand, in the case of
interacting (nonideal) polymers [$v_{pp}(r)\neq 0$], the effective interactions
-- even for monodisperse polymers -- are considerably more complex.  
Although (\ref{omegap1}) then can be formally generalized to
\begin{eqnarray}
\exp(-\beta\Omega_p)&=&\la\exp[-\beta(H_{cp}+H_{pp})]\ra_p \nonumber \\
&=&\sum_{N_p=0}^{\infty}\frac{z_p^{N_p}}{N_p!}
\int{\rm d}{\bf r}^{N_p}\, \prod_{i=1}^{N_c}
\prod_{j=1}^{N_p}(1+f_{ij}^{(c)})\prod_{k<l}^{N_p}(1+f_{kl}^{(p)}),
\label{omegap6}
\end{eqnarray}
where $f_{ij}^{(c)}$ and $f_{ij}^{(p)}$ are the Mayer functions for colloids
and polymers, respectively, practical expressions are less forthcoming.  
Using diagrammatic techniques, Dijkstra {\it et al}~\cite{Dijkstra-pre99}
have further analysed (\ref{omegap6}) and demonstrated its application 
to the phase behaviour of binary hard-sphere mixtures.

\subsection{Systems of ``Soft" Particles}\label{Soft}

Soft particles are here defined as macromolecules having internal 
conformational degrees of freedom.  
Prime examples are flexible polymer chains (linear or branched), 
whose multiple joints allow for many distinct conformations.  While bare
monomer-monomer interactions -- either intrachain or interchain -- can be 
modelled by simple combinations of excluded-volume, van der Waals, and 
Coulomb pair interactions, the total interactions between long, fluctuating 
chains can be highly complex, rendering explicit molecular simulations of 
polymer solutions computationally challenging.

Among several practical techniques developed for averaging over the internal 
structure of soft particles to derive effective pair interactions, we discuss
here two that are specifically suited to polymers in good solvents.
A more extensive survey is given in the review by Likos~\cite{Likos01}.  
One method is based on the general principles of polymer 
scaling theory~\cite{deGennes79}, which describes the properties of polymers 
in the limit of infinite chain length, i.e., segment number $N\to\infty$. 
The starting point is a formal expression for the effective pair interaction 
between the centres of mass, at positions ${\bf R}_1$ and ${\bf R}_2$, 
of two isolated chains (labelled 1 and 2):
\begin{equation}
\beta v_{\rm eff}(R_{12})=-\ln\left(\frac{V^2}{{\cal Z}_1^2}
\int{\cal D}{\bf r}_1\int{\cal D}{\bf r}_2\, \rho_{\rm cm}({\bf R}_1)
\rho_{\rm cm}({\bf R}_2)
\exp[-\beta H[\{{\bf r}_1\},\{{\bf r}_2\}]\right),
\label{veff-polymer}
\end{equation}
where ${\cal Z}_1$ is the partition function of a single isolated chain, 
$\int{\cal D}{\bf r}_{\alpha}$ represents a functional integral over all 
conformational degrees of freedom of chain $\alpha$, 
$\rho_{\rm cm}({\bf R}_{\alpha})$ denotes the number density of the 
centre of mass of chain $\alpha$, and the Hamiltonian $H$ is a functional 
of the conformations, $\{{\bf r}_1\}$ and $\{{\bf r}_2\}$, of the two chains. 
In the special case of two chains, with one end of each chain fixed and the 
two fixed ends separated by a distance $r$, (\ref{veff-polymer}) becomes
\begin{equation}
\beta v_{\rm eff}(r)=-\ln\left(\frac{{\cal Z}_2(r)}{{\cal Z}_2(\infty)}\right),
\label{veff-2chains}
\end{equation}
where ${\cal Z}_2(r)$ is the partition function of the constrained 
two-polymer system and ${\cal Z}_2(\infty)$ is the same in the limit of
infinite separation.
Scaling arguments \cite{Witten-Pincus86} suggest that in the limit 
$N\to\infty$, in which the only relevant length scales are the separation
distance $r$ and the polymer radius of gyration $R_g$, 
${\cal Z}_2(r)/{\cal Z}_2(\infty)\propto (r/R_g)^x$, where $x$ is a 
universal exponent.  It follows that
\begin{equation}
\beta v_{\rm eff}(r)\propto -\ln\left(\frac{r}{R_g}\right), \qquad r\leq R_g,
\label{scaling}
\end{equation}
which describes a gently repulsive effective pair potential.

Similar scaling arguments have been extended to star-branched polymers 
\cite{Likos01,Witten-Pincus86}, consisting of linear polymer chains 
(arms) all joined at one end to a common core.
The same form of effective pair potential results, but with an amplitude
that depends on the number of arms and the solvent quality.  
A refined analysis by Likos {\it et al}~\cite{Likos-stars} leads to an 
explicit expression for the effective pair potential between star polymers,
which extends (\ref{scaling}) to $r>R_g$, specifies the prefactor, 
and is consistent with experimentally measured static structure factors.  

An alternative coarse-graining approach, well-suited to dilute and 
semidilute solutions of polymer coils in good solvents, is based on the 
physically intuitive view of polymer coils as 
``soft colloids"~\cite{Louis-prl00}, whose effective interactions 
may be approximated using integral-equation methods from the 
theory of simple liquids~\cite{HM}.  The basis of this approach is the 
fundamental relation between the pair distribution function $g(r)$, direct 
correlation function $c(r)$, and pair potential $v(r)$ of a simple liquid: 
\begin{equation}
g(r)=\exp[-\beta v(r)+g(r)-c(r)-1-b(r)].
\label{gr}
\end{equation}
Equation~(\ref{gr}) follows directly from the Euler-Lagrange relation
(\ref{rhobr2}) for the nonuniform density of particles $n g(r)$
around a central particle, where the pair potential plays the role of the 
external potential and the bridge function $b(r)$ subsumes all multiparticle 
correlation terms.  Practical implementation begins with a numerical 
calculation of $g(r)$ between the polymer centres of mass, e.g., by 
molecular simulation of explicit self-avoiding random-walk chains, 
and proceeds through inversion of (\ref{gr}) to determine the 
effective centre-to-centre pair potential $v(r)$.  

The inversion of $g(r)$ combines (\ref{gr}) with the Ornstein-Zernike 
relation (\ref{OZ}) and an approximate closure relation for the bridge 
function.  Louis and Bolhuis {\it et al}~\cite{Louis-prl00,Bolhuis-jcp01} 
have demonstrated that the HNC closure, $b(r)=0$, gives an accurate 
approximation for the effective interactions.
The resulting softly repulsive, Gaussian-like, effective pair potential has 
a range comparable to the polymer radius of gyration, an amplitude 
$\simeq 2k_BT$, and is only weakly dependent on concentration, implying the 
relative weakness of effective many-body interactions~\cite{Bolhuis-pre01}.
In self-consistency checks, simulations of simple liquids interacting via
the potential $v(r)$ are found to reproduce, to within statistical errors,
the same centre-to-centre $g(r)$ as simulations of explicit chains.  The same 
method has been applied also to calculate the effective depletion-induced
interaction between hard walls~\cite{Louis-prl00,Bolhuis-jcp01} and between 
colloids in colloid-polymer mixtures~\cite{Bolhuis-prl02,Louis-jcp02}.

The scaling and soft-colloids approaches both reach the common conclusion 
that effective pair interactions between polymers in good solvents are 
ultrasoftly repulsive.  As the centres of mass of two polymers approach 
complete overlap, the pair potential between star polymers diverges very 
slowly (logarithmically), while that between linear chains actually remains 
finite.  These characteristically soft effective interactions contrast 
sharply with the steeply repulsive short-ranged pair interactions between 
colloidal particles, going far in explaining the unique structural 
properties and phase behaviour of polymer solutions.

\section{Summary and Outlook}\label{Outlook}

The key message of this chapter is that soft materials, comprising complex 
mixtures of mesoscopic macromolecules and other microscopic constituents, 
often can be efficiently modelled by preaveraging over some of the degrees 
of freedom to map the multicomponent mixture onto an effective model, with 
fewer components, governed by effective interparticle interactions.  
In general, the effective interactions are many-body in nature and dependent 
on the thermodynamic state of the system.  Briefly surveyed were several 
recently developed statistical mechanical methods, including response theory, 
density-functional theory, and distribution function theory.  These powerful 
methods provide systematic and mutually consistent approaches to approximating 
effective interactions, and have broad relevance to a variety of materials.  
Specific applications were illustrated for electrostatic interactions in 
charged colloids and excluded-volume interactions in colloid-polymer mixtures.

Effective interactions are often simply necessitated by the computational 
impasse presented by fully explicit models of complex systems, especially soft
matter systems with large size and charge asymmetries.  At the same time,
however, effective models can provide conceptual insight that may be difficult
or impossible to extract from explicit models.  Consider, for example, the
subtle interplay of entropy and electrostatic energy in charged colloids or 
the important role of polymer depletion in colloid-polymer mixtures, effects 
that are elegantly and efficiently captured in effective interaction models.
While computational capacity will likely continue to grow exponentially  
in coming years, conceptual understanding of soft materials will also
continue to benefit from the theoretical framework of effective interactions.

\vspace*{0.5cm}
{\large \bf Acknowledgements} \\

Many colleagues and friends have helped to introduce me to the fascinating
world of soft matter physics and the power of effective interactions.  
It is a pleasure to thank, in particular, Neil Ashcroft, J\"urgen Hafner, 
Gerhard Kahl, Christos Likos, Hartmut L\"owen, Matthias Schmidt, and
Alexander Wagner for many enjoyable and inspiring discussions.
Parts of this work were supported by the National Science Foundation 
under Grant No.~DMR-0204020.




%
%

%
%

\end{document}